\documentclass[fleqn,usenatbib]{mnras}
\usepackage[caption=false]{subfig} 
\usepackage{makecell}
\usepackage{graphicx}
\usepackage[version=4]{mhchem}
\graphicspath{{figs/}}
\usepackage{txfonts,float}

\usepackage[T1]{fontenc}
\usepackage{threeparttable} 
\title[The Climate and Compositional Variation of the Highly Eccentric Planet HD 80606 b]{The Climate and Compositional Variation of the Highly Eccentric Planet HD 80606 b -- the rise and fall of carbon monoxide and elemental sulfur}

\author[Shang-Min Tsai et al.]{ 
Shang-Min Tsai,$^{1}$\thanks{E-mail: shang-min.tsai@physics.ox.ac.uk}
Maria Steinrueck,$^{2, 3}$
Vivien Parmentier$^{1, 4}$
Nikole Lewis$^{5}$ \newauthor
Raymond Pierrehumbert$^{1}$
\\
$^{1}$Atmospheric, Oceanic and Planetary Physics, Department of Physics, University of Oxford, Parks Road, OX1 3PU, Oxford, UK\\
$^{2}$Max-Planck Institut f\"ur Astronomie, K\"onigstuhl 17, 69117, Heidelberg, Germany\\
$^{3}$Lunar and Planetary Laboratory, University of Arizona, Tucson, AZ 85719, USA\\
$^{4}$Universit\'e C\^ote d'Azur, Observatoire de la C\^ote d'Azur, CNRS, Laboratoire Lagrange, France\\
$^{5}$Department of Astronomy and Carl Sagan Institute, Cornell University, Ithaca, NY 14853, USA\\
}

\date{Accepted XXX. Received YYY; in original form ZZZ}
\pubyear{2022}

\begin{document}
\label{firstpage}
\pagerange{\pageref{firstpage}--\pageref{lastpage}}
\maketitle
\begin{abstract}
The gas giant HD 80606 b has a highly eccentric orbit (e $\sim$ 0.93). The variation due to the rapid shift of stellar irradiation provides a unique opportunity to probe the physical and chemical timescales and to study the interplay between climate dynamics and atmospheric chemistry. In this work, we present integrated models to study the atmospheric responses and the underlying physical and chemical mechanisms of HD 80606 b. We first run three-dimensional general circulation models (GCMs) to establish the atmospheric thermal and dynamical structures for different atmospheric metallicities and internal heat. Based on the GCM output, we then adopted a 1D time-dependent photochemical model to investigate the compositional variation along the eccentric orbit. The transition of the circulation patterns of HD 80606 b matched the dynamics regimes in previous works. Our photochemical models show that efficient vertical mixing leads to deep quench levels of the major carbon and nitrogen species and the quenching behavior does not change throughout the eccentric orbit. Instead, photolysis is the main driver of the time-dependent chemistry. While \ce{CH4} dominates over CO through most of the orbits, a transient state of [CO]/[\ce{CH4}] $>$ 1 after periastron is confirmed for all metallicity and internal heat cases. The upcoming JWST Cycle 1 GO program will be able to track this real-time \ce{CH4}--CO conversion and infer the chemical timescale. Furthermore, sulfur species initiated by sudden heating and photochemical forcing exhibit both short-term and long-term cycles, opening an interesting avenue for detecting sulfur on exoplanets.


\end{abstract}



%

\section{Introduction}
The orbital eccentricity of planetary systems sheds light on formation history and dynamical evolution. The demographic of eccentricity has been constructed through radial velocity \citep[e.g.,][]{Shen2008}, transit \citep[e.g.,][]{Shen2008} surveys, and synthesis analysis \citep[e.g.,][]{Kipping2013}. The majority of close-in planets have circularized orbits owing to the strong tidal interaction \citep{Pont2011}, which dissipates eccentricity on a timescale that steeply shortens with the semi-major axis at a power-law rate \citep{Goldreich1966}. On the other hand, the discovered exoplanets exhibit a diverse eccentricity distribution beyond a semi-major axis of $\sim$ 0.1 AU \citep{Butler2006,Kane2012}. To understand the mechanisms causing high eccentricity and its impact is of great interest. \cite{Kane2009} have also shown that eccentric planets on average have increasing transit and eclipse probability as compared to their circular-orbit counterparts with the same orbital period, making them ideal observation targets.


The Shoemaker-Levy 9 impact event in 1994 presented a dramatic but one-time example of what can temporal evolution teach us \citep{Harrington2004,Hammel2010}. Individually, an eccentric planet manifests climate variability periodically, as the stellar irradiation varies across the orbit. As opposed to the generally steady state of tidally-locked planets in circular orbits, the eccentricity induced variability provides unique information on how physical and chemical processes interact in the atmosphere. Furthermore, extremely eccentric planets could spend most of the orbit receiving Earth-like irradiation while being briefly heated near periastron and affording direct observations. This climate regime transition provides an illuminating connection between Solar System planets to close-in exoplanets. 


HD 80606 b is a gas giant (4 M$_J$) in the binary system of HD 80606 and HD 80607. The extremely high eccentricity of HD 80606 b (e $\simeq$ 0.933) places it among rare handful of planets with e greater than 0.8. The high eccentricity and close orbit of HD 80606 b are potentially a result of ``Kozai migration'', i.e. the companion star perturbed the planet with an initially inclined orbit into an eccentric orbit, then tidal dissipation drew the planet inward after Kozai oscillation stopped \citep{Wu2003}. During the planet's close encounters, its planetary spin is expected to be quickly synchronized with the orbital revolution at periastron by the strong tidal force (pseudo-synchronous state). However, \textit{Spitzer} observations have suggested that the exact rotational period might differ from the pseudo-synchronous state \citep{deWit2016,Lewis2017}.

\begin{figure}
    \centering
    \includegraphics[width=\columnwidth]{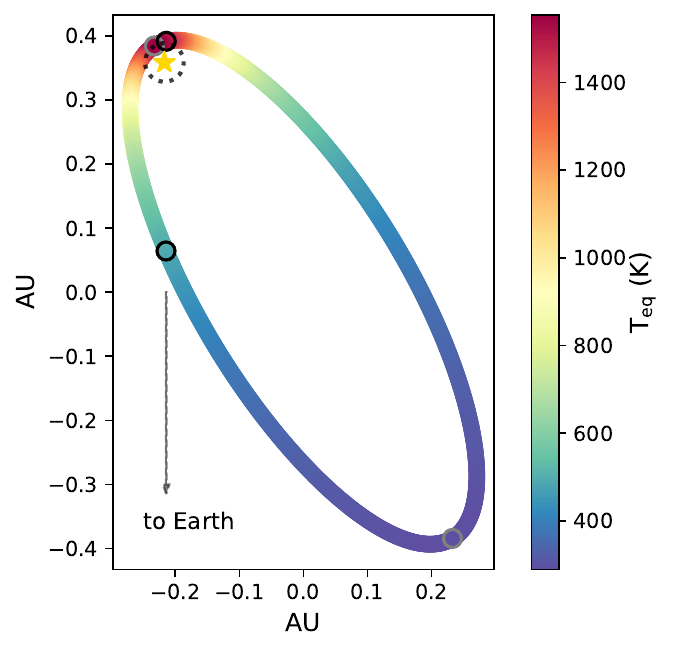}
    \caption{The configuration of the HD 80606 system. The eclipse illustrates the orbit of HD 80606 b, color coded by the planet's equilibrium temperature. The black circles mark the secondary eclipse and primary transit and grey circles mark periastron and apoastron. The dotted circle around the star shows the circular orbit with the same rotational period as that in a pseudo synchronous rotation adopted in our model.}\label{fig:orbit}
\end{figure}

The eccentric configuration and the variation in stellar radiation of the HD 80606 system are illustrated in Figure \ref{fig:orbit}. The unique dynamics of HD 80606 b have attracted various theoretical and observational studies. \cite{Laughlin2009} provided the first look at the \textit{Spitzer} photometric observations at 8 $\mu$m. By measuring the increase of the planetary flux, \cite{Laughlin2009} estimated the global radiative timescale to be $\sim$ 4.5 hours, much shorter than that in cooler atmospheres in Solar System ($\sim$ days). \cite{deWit2016} followed up with the 4.5 $\mu$m channel observation spanning a longer phase coverage around the periastron. \cite{deWit2016} further considered the effect of the revolving substellar longitude by planetary rotation and inferred a rotational period longer than the pseudo-synchronous period ($\sim$ 93 hours). The tidal quality factor of HD 80606 b is also estimated to be much higher than that of Jupiter, indicating a lower rate of tidal dissipation that is consistent with the high eccentricity the system retained. Finally and excitingly, HD 80606 b will soon be observed by the NIRSpec and MIRI instruments on JWST during Cycle 1 General Observers (GO) program JWST-GO-2008 (PI Kataria) and JWST-GO-2488 (PI James Sikora). These observations will provide time-series spectra to reveal how the planet responds to the flash heat around periastron in detail. 

From the modeling effort, 1D time-stepping radiative models have been applied to studying the thermal evolution of the full phase \citep{Iro2010,Mayorga2021}. To probe the atmospheric dynamics, \cite{Langton2008} applied a sallow-water model and identified the fast zonal flow driven by rapid heating near periastron. \cite{Lewis2017} explored the effects of rotational period with a 3D GCM and found that the timing of the planetary flux strongly depends on rotation. 
\cite{Lewis2017} also suggested that observed phase curve near periastron could be explained by the emergence and dissipation of optically thick clouds.  

While most of the previous works about eccentric planets focused on the thermal response, they often assume a constant composition when analyzing the light curves. However, changes in the atmospheric chemistry can significantly alter the photospheric levels probed by the observations \citep{Dobbs-Dixon2017,Parmentier2021} and hence the derived planetary properties (e.g., rotational rate). The assumption of maintaining thermochemical equilibrium instantaneously is also not expected to hold due to various disequilibrium processes, such as atmospheric mixing, photochemistry, and the stellar flux variation along the eccentric orbit, etc. \cite{Visscher2012} provides an analytical overview of the chemical conversion on eccentric planets and indicates that the disequilibrium processes play an important role in the CO--\ce{CH4} interconversion.   


Despite the great effort devoted to study HD 80606 b, a comprehensive understanding of how chemistry and dynamics evolve in an eccentric system is still lacking. To unveil the essential atmospheric properties en masse, we assemble 3D GCM and photochemical models in this work to study the climate and compositional response in tandem. In Section \ref{sec:3D}, we discuss GCM modeling and the global thermal and dynamical responses to the rapid heating. In Section \ref{sec:1D}, we introduce the 1D time-dependent photochemical model and show the chemical response from simple timescale comparison to detailed photochemical pathway analyses. We bring together the modeling results and present synthetic spectra at the end of \ref{sec:1D}. Opportunities for future work and observations are addressed in Section \ref{sec:discussion} and we sum up the highlight of this study in Section \ref{sec:sum}. 


\section{The thermal and dynamical variations}\label{sec:3D}
\subsection{The setup of 3D GCM simulations}\label{sec:GCM}
We simulate the 3D global thermal and dynamical properties of HD 80606 b's atmosphere with the SPARC/MITgcm model \citep{Showman2009,Adcroft2004}. The SPARC/MITgcm model has been extensively applied to study the circulation of gas giants and brown dwarfs \citep[e.g.][]{Showman2009,Parmentier2016,Powell2019,Steinrueck2019,Tan2019}. For radiative transfer, we assume chemical equilibrium and use the correlated-k method with 8 Gauss points within each of the 11 wavelength bins. A solar spectrum is used as an analog star for HD 80606. The simulations employ a resolution of C32 on the cubed-sphere grid and 53 vertical layers over 170 and 2.4$\times 10 ^{-6}$ bar. While the rotation period is one of the fundamental planetary properties that control the climate dynamics, it is challenging to place tight constraints on the rotation state of HD 80606 b. In this work, we follow \cite{Hut1981,Hut1982} and assume that the planet is ``pseudo-synchronized'' with its host star, i.e., the planet's rotation period is equal to the revolution period of a circular orbit at periastron, which is the dynamically stable configuration. The pseudo-synchronous rotation period is determined by the orbital period and eccentricity \citep{Hut1981}
\begin{equation}
\tau_{\textrm{rot}}=\tau_{\textrm{orb}} \times \frac{(1+3e^2+\frac{3}{8}e^4)(1-e^2)^{3/2}}{1+\frac{15}{2}e^2+\frac{45}{8}e^4+\frac{5}{16}e^6}.
\end{equation}
We adopted a pseudo-synchronous rotational period of 40.475 hours for HD 80606 b. \cite{deWit2016} estimated the rotational period of 93 hours from the thermal fluxes observed by Spitzer, about twice as long as that of pseudo-synchronous rotation though. We defer the readers to \cite{Lewis2017} for the dynamical effects of different rotational periods for HD 80606 b.


\cite{Liu2018} performed high-resolution spectral analysis and found the metallicity of HD 80606 to be about twice the solar value. Additionally, tidal dissipation is expected to increase the internal heating of eccentric planets \citep{Leconte2010}. Therefore, we further explore the effects of atmospheric metallicity and internal heating. We performed models for 1 time, 5 times solar metallicity, and internal temperature $T_{\textrm{int}}$ = 100 , 400 K, respectively. A summary of the model parameters used in this study are listed in Table \ref{table:gcm_para}. A periodic steady state is found after just a few orbits, where the temperature and wind structures at the same orbital position between different orbits show negligible changes. The output after nine orbits of simulation runs is used in this work.


\begin{table} 
\begin{threeparttable} 
\caption{Model parameters of HD 80606 b for the SPARC/MITgcm.}             
\label{table:gcm_para}      
\centering          
\begin{tabular}{c c c}     
\hline\hline
symbol & description & value\\
\hline 
R$_p$ & planet radius & 0.98 R$_\textrm{J}$\tnote{a}\\ 
g & gravity & 10510 cm$^2$/s\tnote{a}\\
a & semi-major axis & 0.449 au\tnote{a}\\
e & eccentricity & 0.933\tnote{a}\\
$\Omega$ & rotational period & 40.475 hr\\
c$_P$ & heat capacity & 13000 J kg$^{-1}$ K$^{-1}$ \\
R & specific gas constant & 3714 J kg$^{-1}$ K$^{-1}$ \\
T$_{\textrm{irr}}$ & \makecell{irradiation temperature\tnote{b}\\at periastron, apoastron} &  415, 2230 K\\
T$_{\textrm{int}}$ & internal temperature & 100, 400 K\\
 & Resolution & C32\\
& Simulated time & 9.5 orbits (1059 Earth days)\\ 
\hline                  
\end{tabular}
\begin{tablenotes}\footnotesize
\item [a] \cite{Pont2009}
\item [b] T$_{\textrm{irr}}$ = T$_\star$ (R$_\star$/D)$^{1/2}$ where R$_\star$ is the stellar effective temperature and D is the instantaneous orbital distance.
\end{tablenotes}
\end{threeparttable}
\end{table}

\subsection{The rise of temperature and winds around periastron}
\begin{figure*}
    \centering
    \includegraphics[width=\columnwidth]{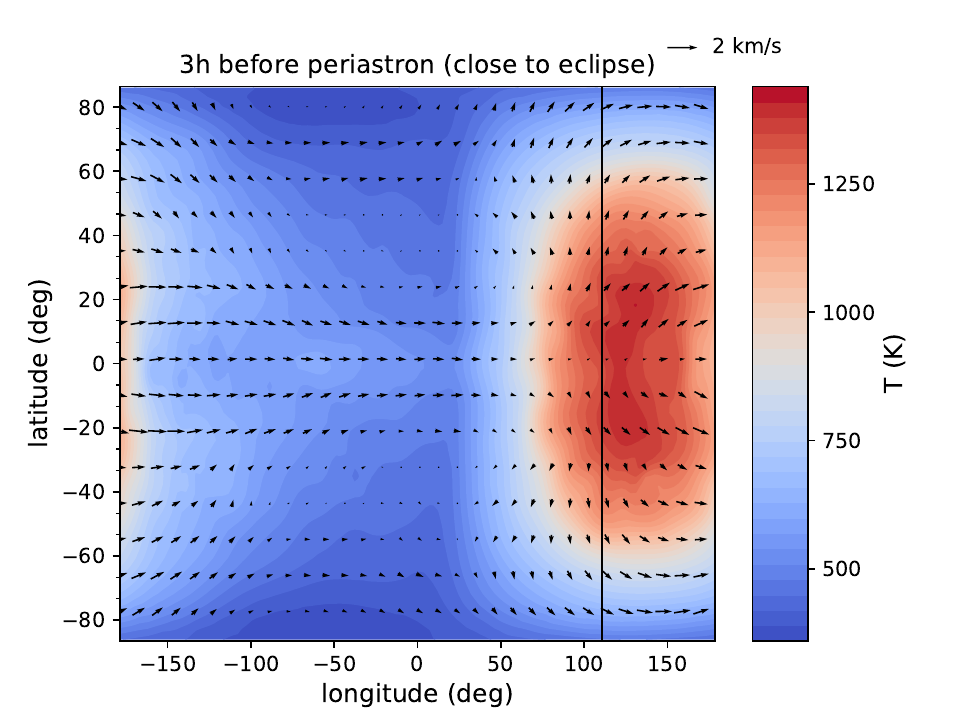}
    \includegraphics[width=\columnwidth]{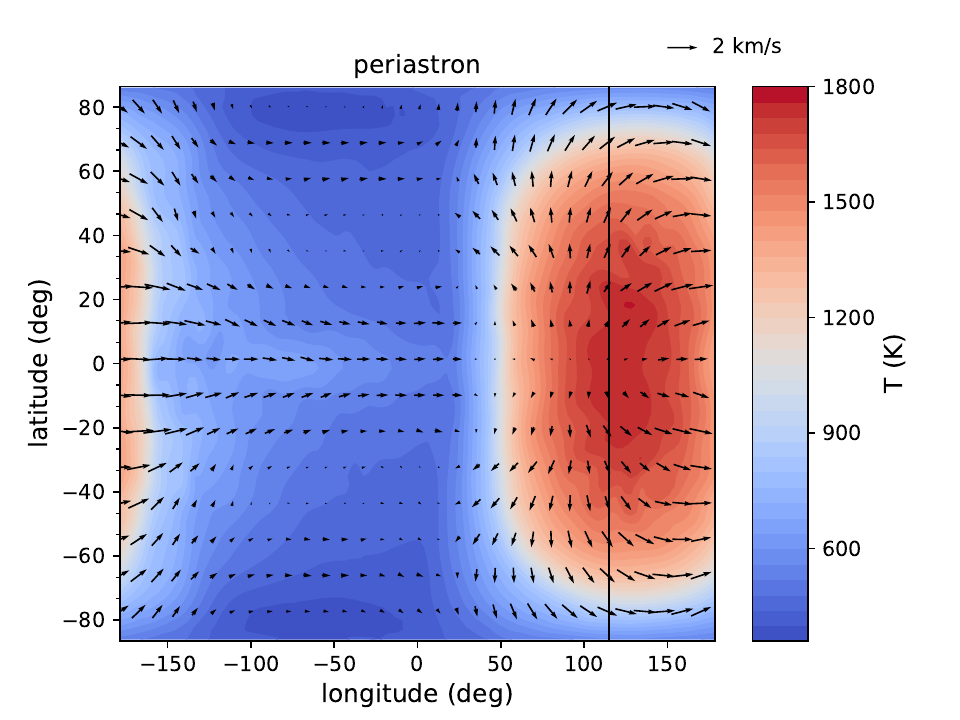}
    \includegraphics[width=\columnwidth]{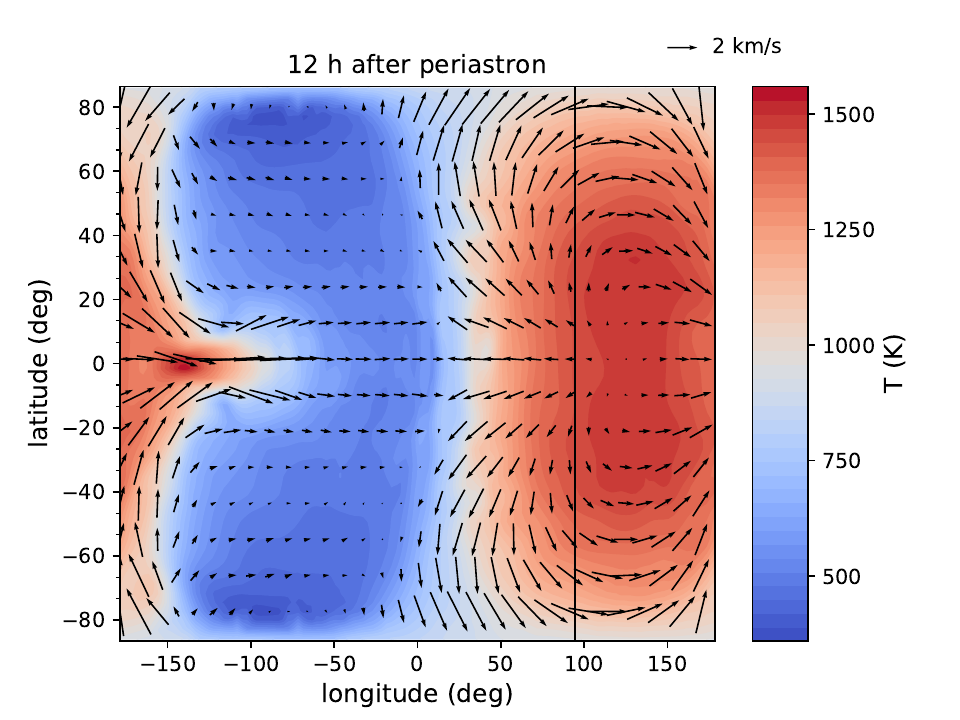}
    \includegraphics[width=\columnwidth]{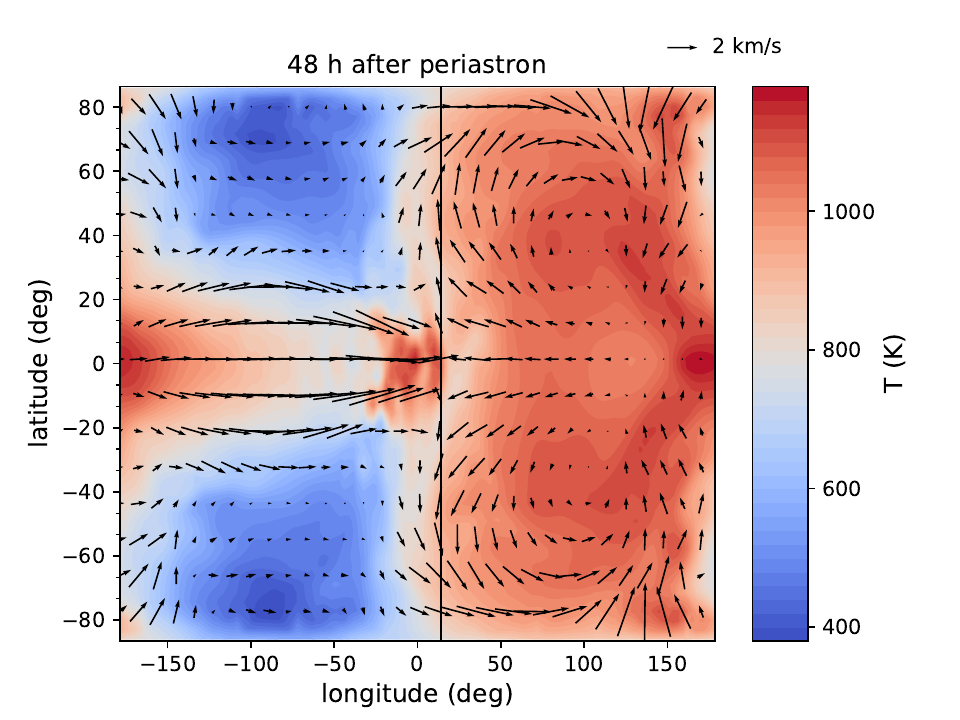}
    \includegraphics[width=\columnwidth]{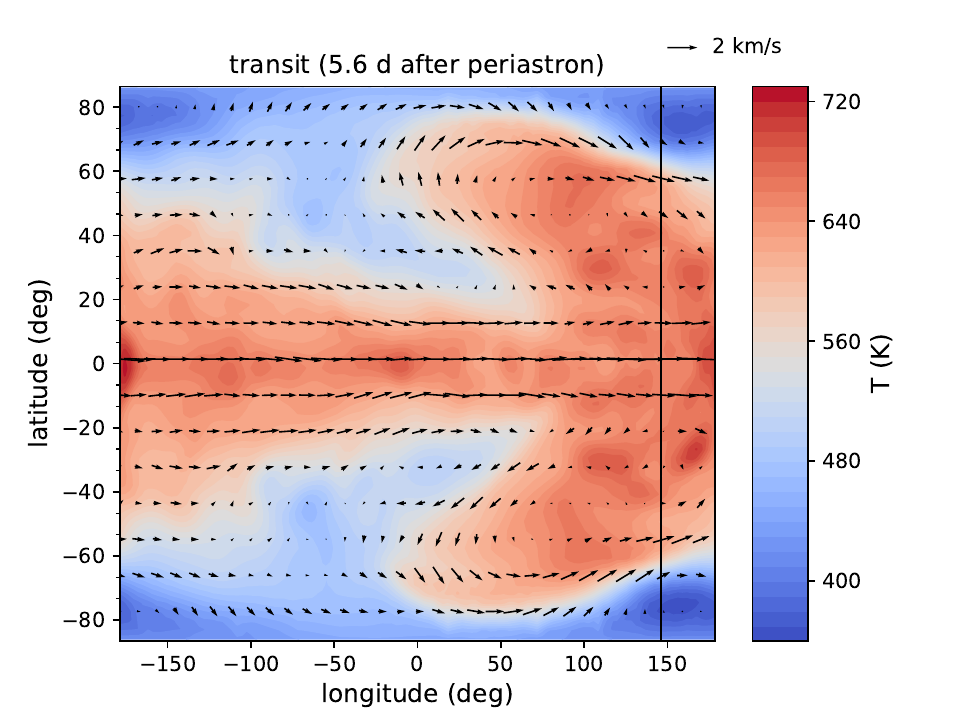}
    \includegraphics[width=\columnwidth]{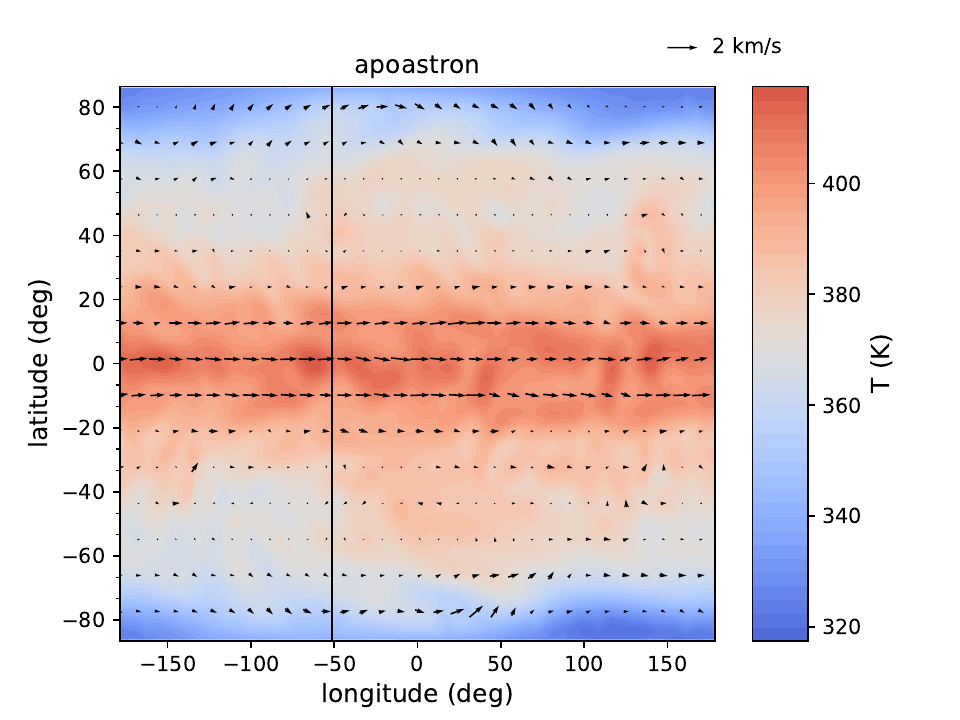}
    \caption{Temperature (color contours) and horizontal winds (arrows) at 160 mbar level of our HD 80606 b GCM with solar metallicity and $T_{\textrm{int}}$ = 100 K (nominal model). Each panel displays a snapshot at different times and the black vertical line indicates the longitude where the star is directly overhead (varying in the pseudo synchronous rotation). The wind vectors scale the same but the temperature colors do not.}\label{fig:GCM_time} 
\end{figure*}

Owing to the high eccentricity of HD 80606 b, the stellar irradiation received by the planet shifts drastically from that similar to Earth at apoastron to about 800 times higher at periastron, raising the equilibrium temperature from $\sim$ 300 K to $\sim$ 1500 K (Figure \ref{fig:orbit}). The planet spends more than half of the orbital period with a distance greater than 0.7 AU to its host star and only about two days with a distance less than 0.1 AU. We can in fact regard HD 80606 b as a warm Jupiter that enters a brief period of hot Jupiter regime every $\sim$ 100 days.

The thermal and dynamical responses to the rapid heating near periastron are illustrated in Figure \ref{fig:GCM_time}. The pulse heating excites planetary-scale waves that associate with the momentum transport of the mean flow \citep{Showman2011,Tsai2014,Showman2020}. The patterns resemble those commonly seen in the spin-up stage for circulation under stationary forcing \citep{Arnold2012,Debras2020,Hammond2020}. Typically, the zonal jet on tidally locked planets takes about 10--100 Earth days to approach equilibrium from the initial rest state \citep[i.e. the spin-up process;][]{Showman2011,Hammond2018,Hammond2020}. Hence, the rapid heating of $\sim$ 2 days near periastron does not provide enough time for the jet to fully develop before the forcing fades out. The jet speed evolves steadily throughout the orbit between 500 m/s and 2000 m/s but the circulation pattern outside of the periastron passage does not vary much and is similar to that shown at apoastron.

We can gain physical insight into the transition of dynamical regimes by comparing the circulation of HD 80606 b at different orbital phases with the grid of models with different stellar fluxes and rotational periods in \cite{Showman2015}. The pseudo-synchronous rotational period of HD 80606 b is close to the median rotation case ($\Omega_{\textrm{med}}$; hereafter following the notation in \cite{Showman2015}) in \cite{Showman2015}. The eccentric planet swings between the hot state ($\textrm{H}\Omega_{\textrm{med}}$) and the cold state ($\textrm{C}\Omega_{\textrm{med}}$). \cite{Showman2015} find that the hot state is dictated by the day-night thermal forcing while the cold state exhibits zonal symmetry and is mainly driven by rotation effects. 

\begin{figure}
    \centering
    \includegraphics[width=\columnwidth]{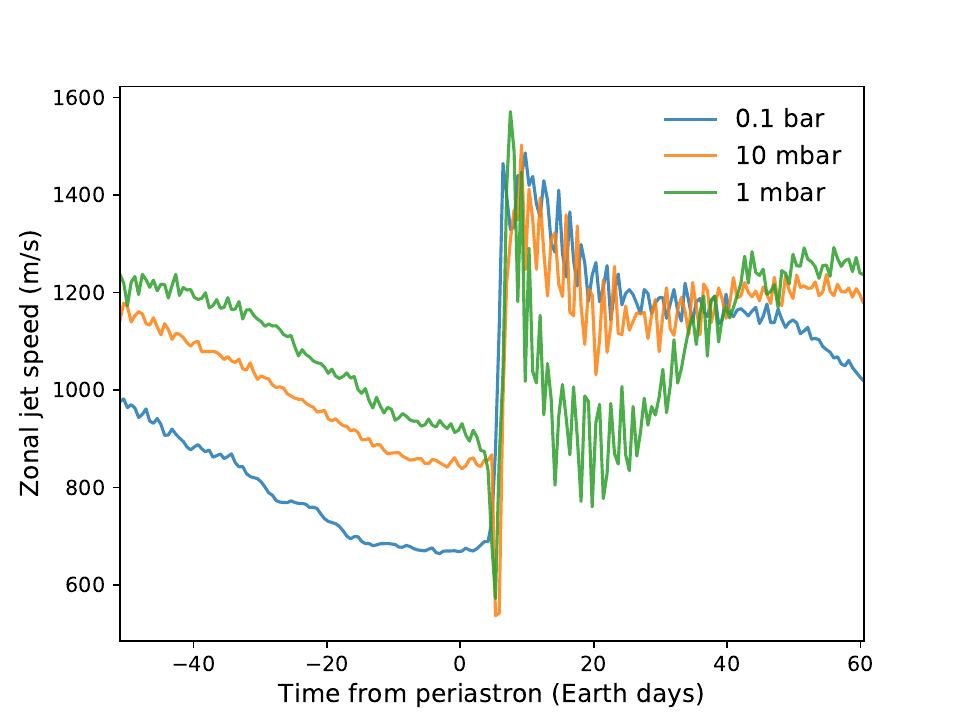}
    \caption{The evolution of the zonal jet speed at different pressure levels averaged over $\pm$ 20 latitudes for an orbit.}\label{fig:U-orbit} 
\end{figure}

The high eccentricity makes HD 80606 b stay in $\textrm{H}\Omega_{\textrm{med}}$ only for less than two days before returning to $\textrm{C}\Omega_{\textrm{med}}$. For most of the orbit, it shares circulation features similar to $\textrm{C}\Omega_{\textrm{med}}$, except the equatorial wind speed is faster than that in $\textrm{C}\Omega_{\textrm{med}}$. In this regime, the planet received weaker irradiation with the latitudinal temperature gradient predominating over the day-night temperature difference. The stellar irradiation swiftly increased by two orders of magnitude as the planet approaches periastron and enters $\textrm{H}\Omega_{\textrm{med}}$ regime. In this regime, the thermal and wave structures resemble those in $\textrm{H}\Omega_{\textrm{med}}$ but without the fast equatorial jet. This feature can be understood by considering the radiative timescale of an atmosphere with characteristic pressure $P$ and equilibrium temperature $T_{\textrm{eq}}$  \citep[e.g.][]{showman2002,Mayorga2021},
\begin{equation}
\tau_{\textrm{rad}} \sim \frac{P c_P}{4 g \sigma T_{\textrm{eq}}} \sim 1 \times 10^3 (\frac{1500 \textrm{K}}{T_{\textrm{eq}}})^3 \textrm{sec}     
\end{equation}
where $c_P$, $g$, and $\sigma$ are heat capacity, gravity, and the Stefan-Boltzmann constant, respectively. $\tau_{\textrm{rad}}$ of HD 80606 b is in the order of hours near periastron, which is consistent with the estimate from the phase curve observation \citep{Laughlin2009,deWit2016}. Therefore, the thermal response of the atmosphere ($\sim$hours) is much faster than the growth time of the jet ($\sim$10--100 days). This thermal forcing initiates the Matsuno-Gill type \citep{Matsuno1966,Gill1980,Showman2011} standing waves that reset the mean flow but at the same time
pump momentum toward the equator. In the end, although the flash forcing does not allow the jet to develop during the short periastron passage, it contributes to accelerating the equatorial wind over time out of the periastron passage. This explains the stronger ($\sim$ 1000 m/s faster) equatorial jet of HD 80606 b outside the periastron passage, while comparing to $\textrm{C}\Omega_{\textrm{med}}$ in \cite{Showman2015} without the ``flash forcing''. The mean equatorial jet speed over a period is shown in Figure \ref{fig:U-orbit}, where the sinusoidal cycle associates with wave-pumping acceleration for about half of the orbit and gradually slowing down of the jet for the second half. 

\begin{figure*}
   \centering
   \includegraphics[width=1.75\columnwidth]{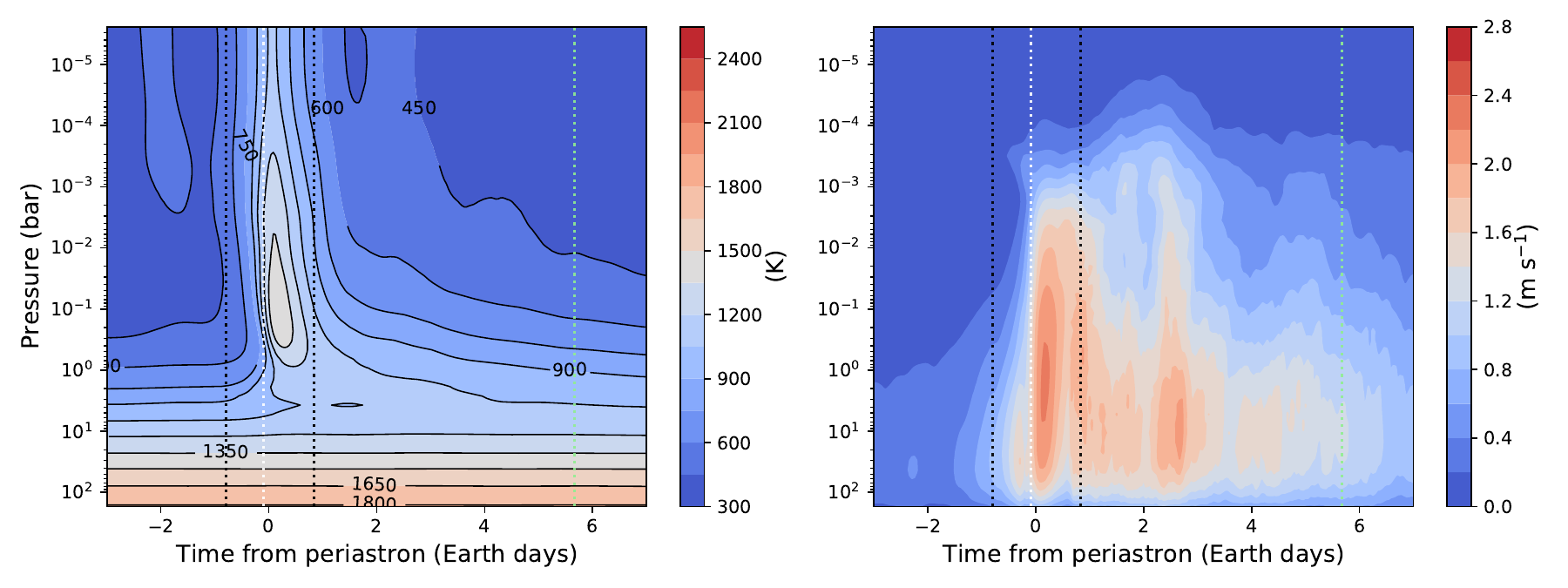}
   \includegraphics[width=1.75\columnwidth]{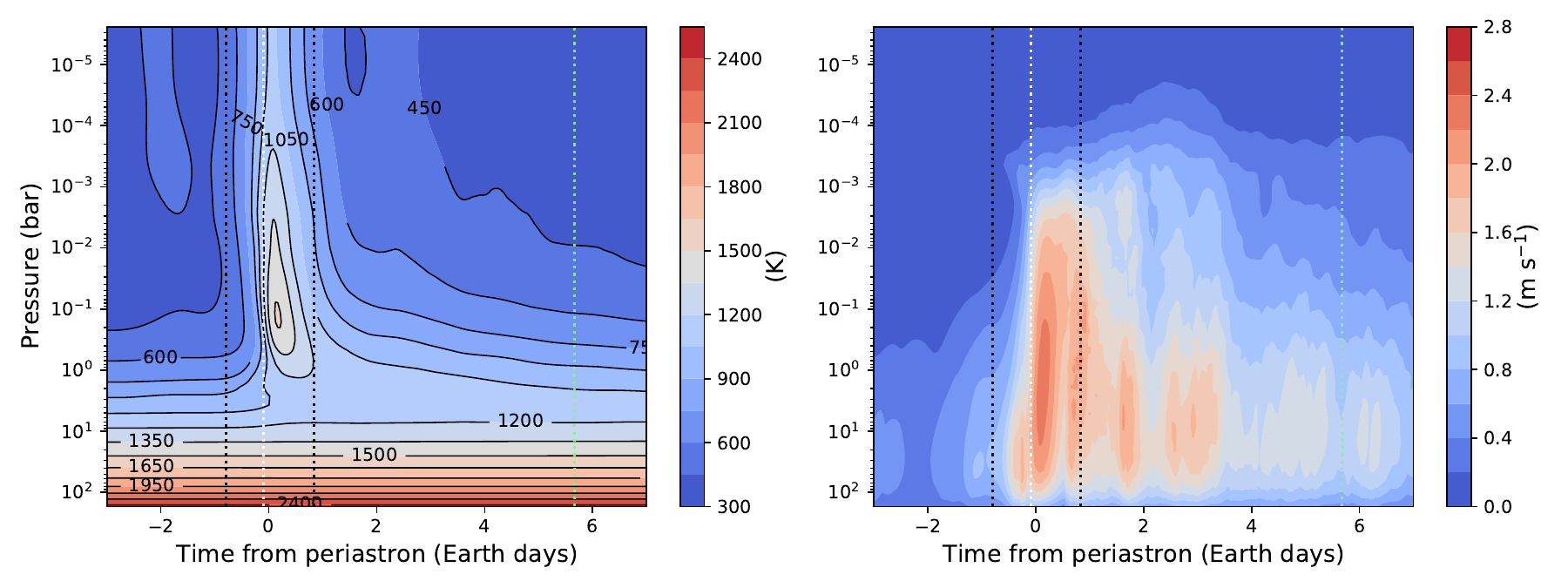}
   \includegraphics[width=1.75\columnwidth]{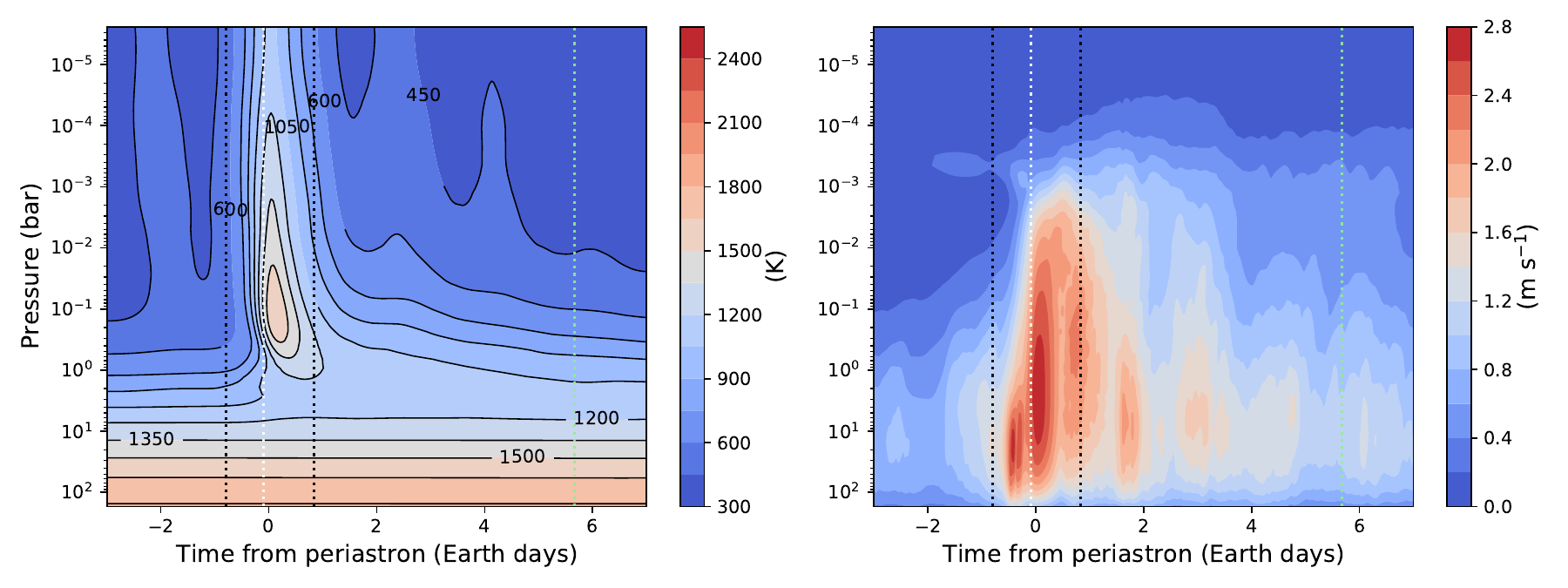}
   \includegraphics[width=1.75\columnwidth]{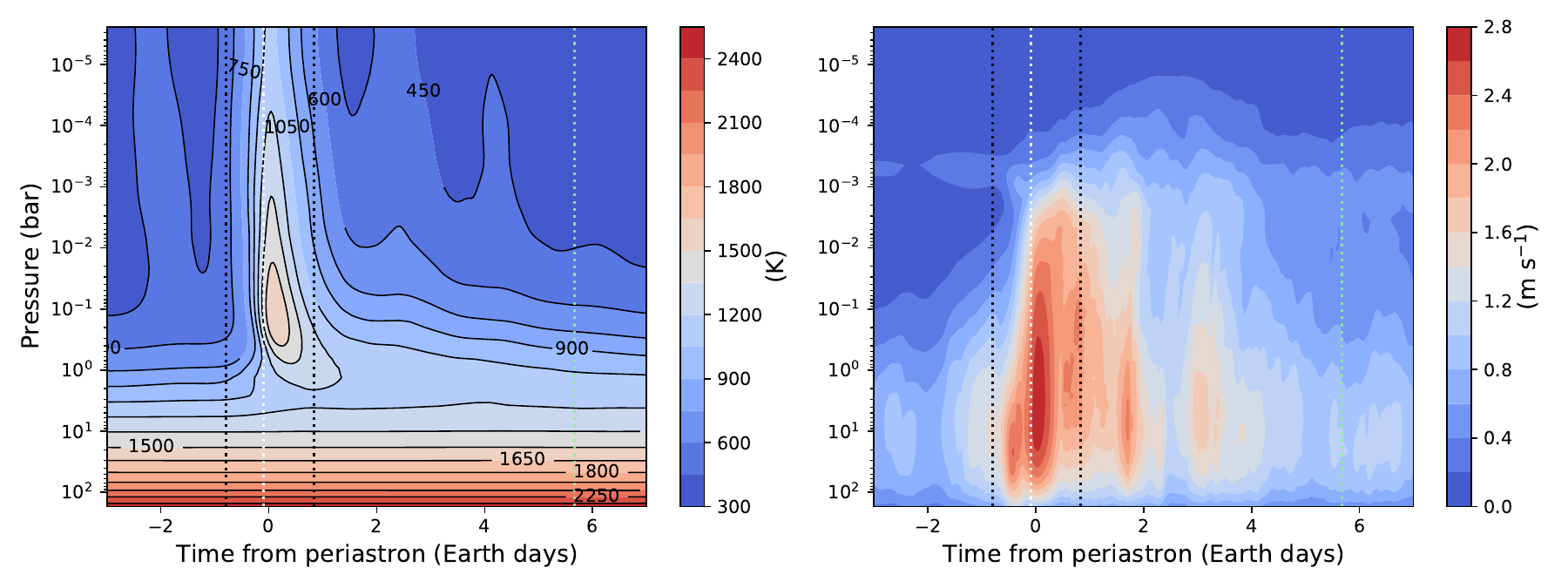}
      \caption{Dayside averaged temperature (left) and root-mean-square vertical velocity (right) at different time relative to periastron from the 3D general circulation models of HD 80606 b, assuming solar metallicity and $T_{\textrm{int}}$ = 100 K (top tow), solar metallicity and $T_{\textrm{int}}$ = 400 K (second row), $5 \times$ solar metallicity and $T_{\textrm{int}}$ = 100 K (third row), and $5 \times$ solar metallicity and $T_{\textrm{int}}$ = 400 K (bottom row). The white dotted line indicates secondary eclipse ($\sim$ 2 hours before periastron) and the lime dotted line denotes the primary transit. The black dotted lines enclose the period where the planet is in synchronous rotation. The same lines are used in Figure \ref{fig:peri-contour}, \ref{fig:peri-contour-noOrbit}, and \ref{fig:peri-contour-S} as well.}
         \label{fig:TPs}
\end{figure*}

Figure \ref{fig:TPs} presents the overall response near periastron for different metallicities and internal temperatures, showing the dayside averaged temperature and root-mean-square (RMS) vertical wind as a function of time relative to periastron. A transient thermal inversion occurs  immediately after periastron in all cases. Both temperature and vertical wind remain elevated until transit ($\sim$ 5.6 days after periastron) with respect to those before periastron. Compared to the model with solar metallicity and $T_{\textrm{int}}$ = 100 K (referred as the nominal model hereafter), we find that a higher metallicity increases the atmospheric opacity and enhances the shortwave absorption, leading to a warmer stratosphere and a more pronounced updraft near periastron. On the other hand, the temperature and vertical wind in altitudes above the 1-bar level are not directly sensitive to internal heat.

\section{The chemical variations}\label{sec:1D}
\subsection{The setup of 1D time-dependent photochemical model} 
We then simulate the time-dependent atmospheric composition for HD 80606 b based on the temperature and wind structures from the GCM. We employ a well-vetted 1D photochemical kinetics model VULCAN \citep{tsai17,Tsai2021}. The photochemical model treats thermochemistry, photochemistry, diffusion, and condensation processes. We apply the C–H–N–O–S chemical scheme \footnote{\url{https://github.com/exoclime/VULCAN/blob/master/thermo/SNCHO_photo_network.txt}}, including 90 species with 1028 thermochemical reactions and 60 photodissociation reactions. We adapted VULCAN to have temperature and stellar irradiation that continuously varied with time according to the orbital position of eccentric planets. The time series of GCM output provides the dayside average properties as input for VULCAN to simulate the compositional variation. HD 80606 is a G5 star with effective temperature of 5645 K and we used the solar spectrum \citep{Gueymard2018} as an analogue. Since the planet is assumed to be in pseudo-synchronous rotation, we define the dayside at periastron to be the dayside hemisphere throughout the orbit, i.e. our 1D photochemical model represents the dayside that is quasi-synchronous around periastron and keeps track of the same geographical hemisphere for the rest of the orbit. In this way, the dayside model captures the shifts of stellar irradiation near periastron, while the diurnal cycle of instellation makes the dayside and global average temperatures and wind almost identical away from periastron. We do not include the compositional feedback to the 3D GCM self-consistently and will discuss the limit of our 1D chemical model in \ref{sec:disscusion}. The same planetary parameters in Table \ref{table:gcm_para} and a zenith angle of 58 $^{\circ}$ are used in the dayside-average photochemical model. 

\begin{figure}
    \centering 
    \includegraphics[width=\columnwidth]{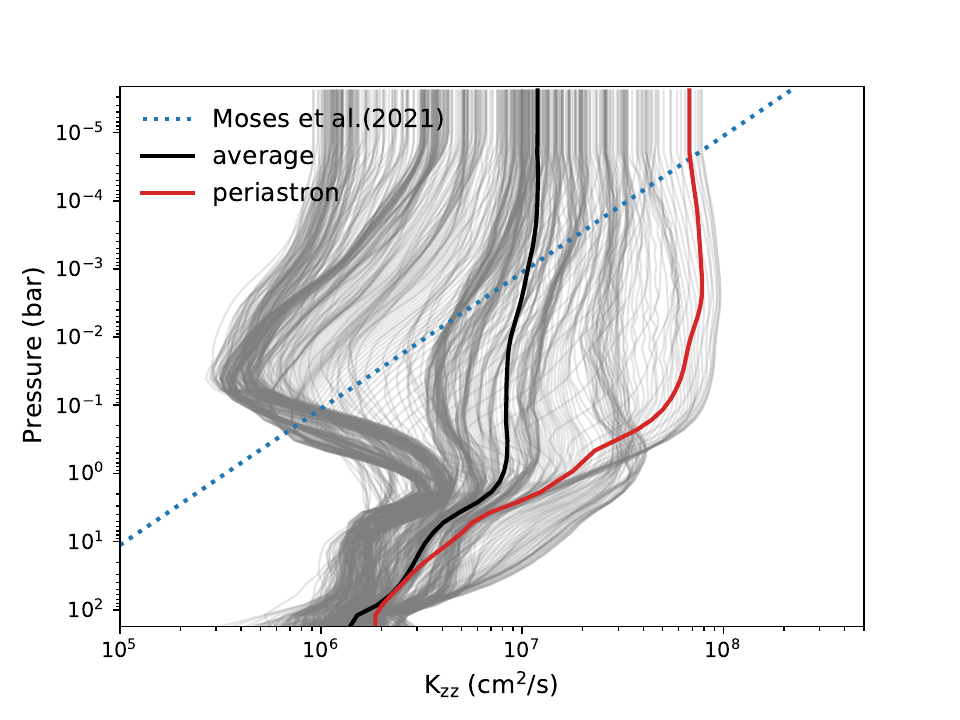}
    \caption{The eddy diffusion coefficient profiles derived from the RMS vertical winds of the GCM at different time along the orbit (grey lines), with the orbital averaged profile shown in black and that at periastron in red. For comparison, the blue dashed line show the $K_{\textrm{zz}}$ profiles from the scaling relation \citep{Moses2021} with $T_{\textrm{eq}}$ calculated for an orbital distance of semi-major axis.}\label{fig:kzz_orbit} 
\end{figure}

 
Atmospheric mixing is commonly parameterized by eddy diffusion in 1D models. To track the variation of vertical mixing with time, we derive the eddy diffusion coefficients from the dayside averaged vertical wind ($w_{\textrm{rms}}$) at a given time according to the mixing length theory $K_{\textrm{zz}}$ = 0.1 $w_{\textrm{rms}} \times H$, with $H$ being the pressure scale height and the factor of 0.1 following the empirical factor from tracer studies \citep[e.g.][]{Parmentier2013,Charnay2015}. The temperatures and $K_{\textrm{zz}}$ are kept at the same values from those at the top boundary of the GCM ($\sim$10$^{-5}$ bar ) when extending to $10^{-8}$ bar for the photochemical model. Figure \ref{fig:kzz_orbit} shows the eddy diffusion profiles around an orbit, compared to the temperature scaling expression in \cite{Moses2021}. Our wind-derived $K_{\textrm{zz}}$ shows a weaker pressure dependence, but the magnitude is consistent with the bounds given by the expression in \cite{Moses2021}. We further explore the sensitivity to eddy diffusion in Section \ref{sec:Kzz}. 


In this time-resolved photochemical model, one orbit of HD 80606 b is divided into 574 temporal grid points, with a separation of 1 hour near periastron to resolve the rapid variation and 13.3 hours for the rest of the orbit. The temperature and eddy-diffusion at each grid are updated from the GCM output and the UV flux is adjusted according to the orbital distance. The stepsize in the photochemical model is bounded by the grid size (e.g. 1 hour near periastron) to resolve the orbital evolution. To reduce the convergence time, we first run the chemical model at apoastron to the steady state. Then the steady-state composition at apoastron is used to initialize the time-dependent model with orbital motion. With our initialization and the planet's long period (about 111 days), we find it takes less than 5 orbits to reach a periodic steady state. We have also tested the temperature dependence of UV cross sections of \ce{CO2} and \ce{H2O} \citep{Tsai2021} and found negligible differences. After obtaining the composition, we use the open-source radiative-transfer tools HELIOS \citep{Malik2019} to generate emission spectra and PLATON \citep{Zhang2019,Zhang2020} for transmission spectra in Section \ref{sec:spectra}.


\subsection{The variation of chemical composition} 
\subsubsection{Orbital-position independence of vertical quenching}\label{sec:st_quench}
\begin{figure}
   \centering
   \includegraphics[width=\columnwidth]{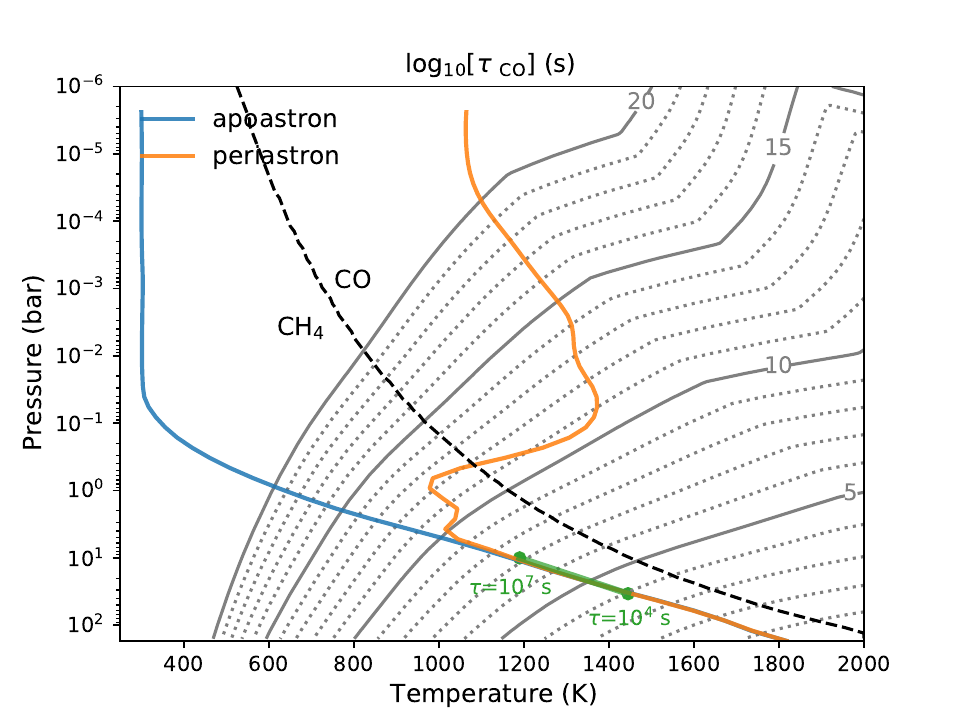}
   \includegraphics[width=\columnwidth]{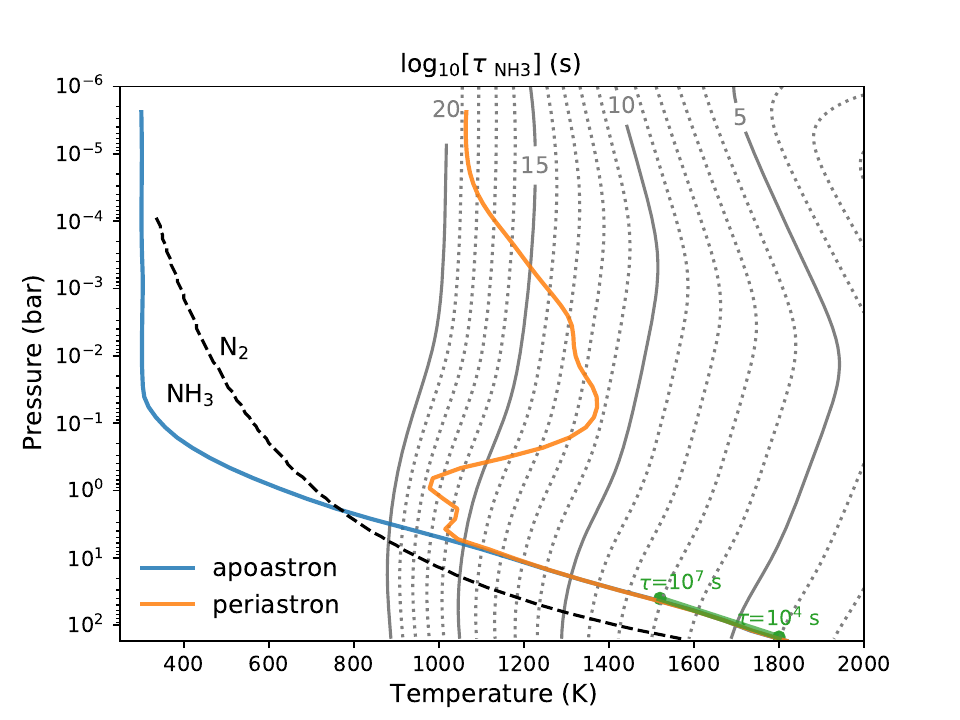}
    \caption{The temperature profiles of our HD 80606 b model at apoastron (blue) and periastron (orange) overlaying the contours of the chemical timescales of CO and \ce{NH3}, given by the expressions in \protect\cite{tsai18}. The dashed line shows where \ce{CH4} and CO (\ce{NH3} and \ce{N2}) are in equal abundance. The green circles indicate the quench levels corresponding to vertical mixing timescales of 10$^4$ -- 10$^7$ s taking into account mixing-length uncertainty (estimated from $\tau_{\textrm{mix}} \sim H^2/ K_{\textrm{zz}}$ in the atmosphere below 1 bar). The quench levels are deep such that the temperature at the quench levels remains unchanged along the orbit.}\label{fig:timescale}
\end{figure}

First, following \cite{Visscher2012}, we will rely on equilibrium chemistry and timescale comparison to gain insight. Figure \ref{fig:timescale} illustrates the temperature profiles and the transitions of \ce{CH4}--CO and \ce{NH3}--\ce{N2}, overlying the chemical timescales of CO and \ce{NH3}. Thermochemical equilibrium predicts \ce{CH4} dominates CO for the entire atmosphere at cold apoastron, while CO can take over \ce{CH4} at altitudes above the 1-bar level at periastron as the temperature rapidly increased. Similarly, \ce{NH3} is favoured over \ce{N2} at apoastron between 1 bar and 0.1 mbar, while \ce{N2} predominates at periastron. In reality, however, as the timescale of chemical conversion exceeds that of atmospheric transport at lower temperatures/pressures, the species retain uniform distributions above the levels where the two timescales are equal, i.e. 
the quench levels \citep[e.g.,][]{Visscher2010,tsai18}. The quenched abundances generally provide the basis distribution in the observable region \citep{Moses2014,Baxter2021,Fortney2020}. 

For eccentric planets, one might expect that the shift of temperatures and the strength of mixing would lead to temporal variations in the quenching process. Yet it is not the case for HD 80606 b. Figure \ref{fig:timescale} indicates that both CO and \ce{NH3} are quenched in the deep atmosphere between 10 and 100 bar, where the radiative time scale is too long to manifest any temperature variations along the orbit. Hence, for a given $K_{\textrm{zz}}$, the quench levels are independent of the thermal variations throughout the orbit. The elevated vertical wind (Figure \ref{fig:GCM_time}) after periastron also turns out to be not important for affecting the quench levels, because the steep decline of chemical timescales implies that the quench levels are not sensitive to merely a few factors of variation of $K_{\textrm{zz}}$. Ultimately, we find the quenched abundances of the major carbon and nitrogen species remain unchanged throughout the orbit. 
In the next section, we will discuss how the chemical response is predominantly driven by photochemistry. 



\begin{figure*}
   \centering
    \includegraphics[width=\columnwidth]{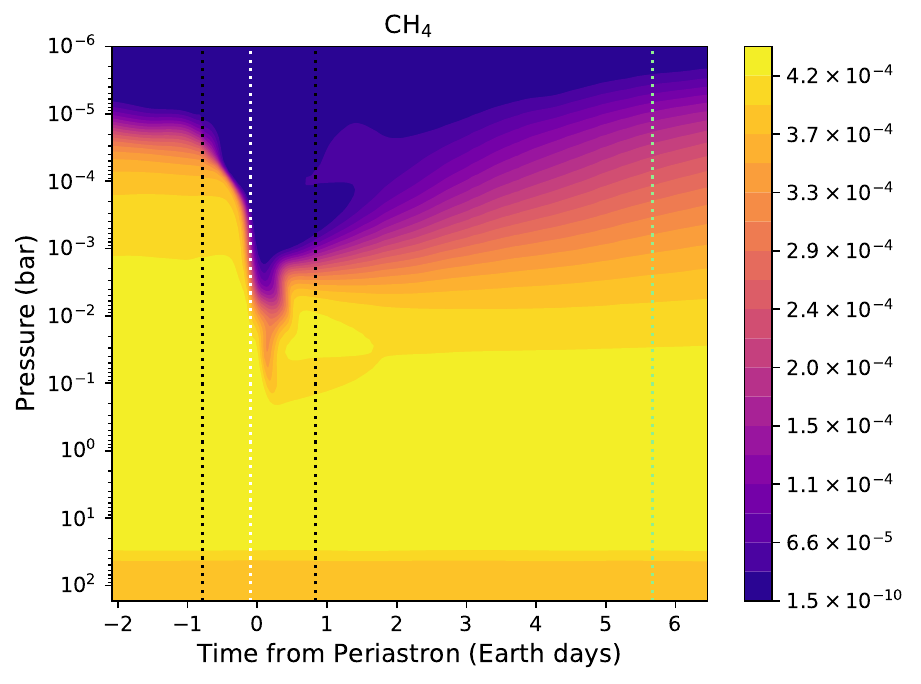}
    \includegraphics[width=\columnwidth]{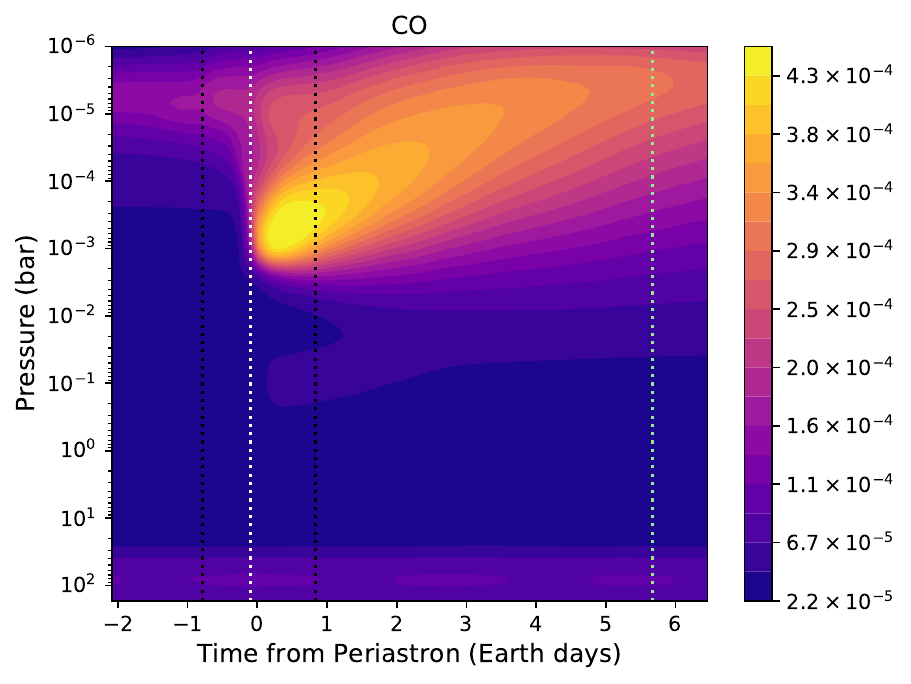}
    \includegraphics[width=\columnwidth]{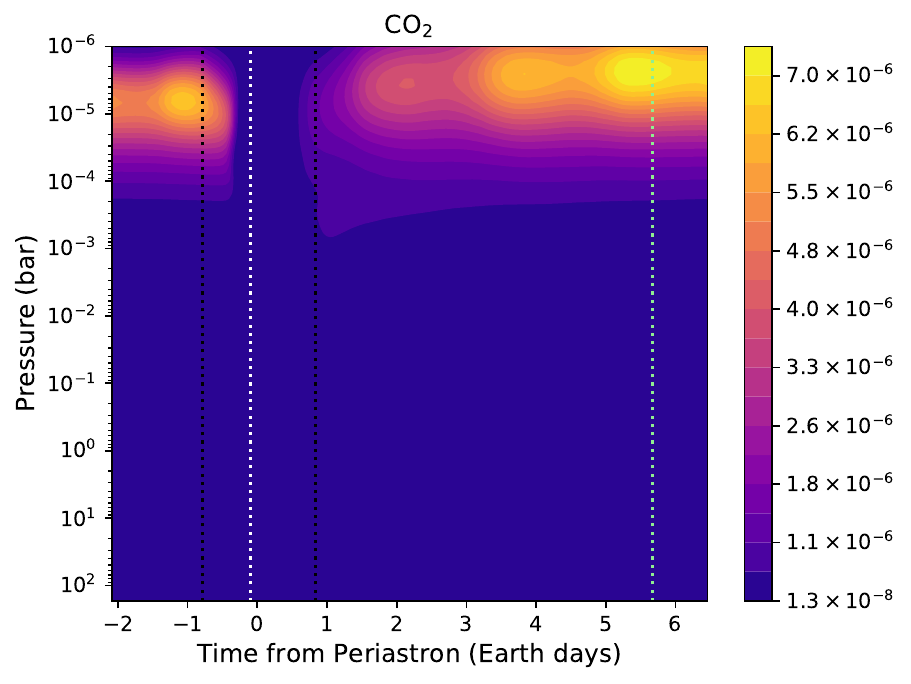}
    \includegraphics[width=\columnwidth]{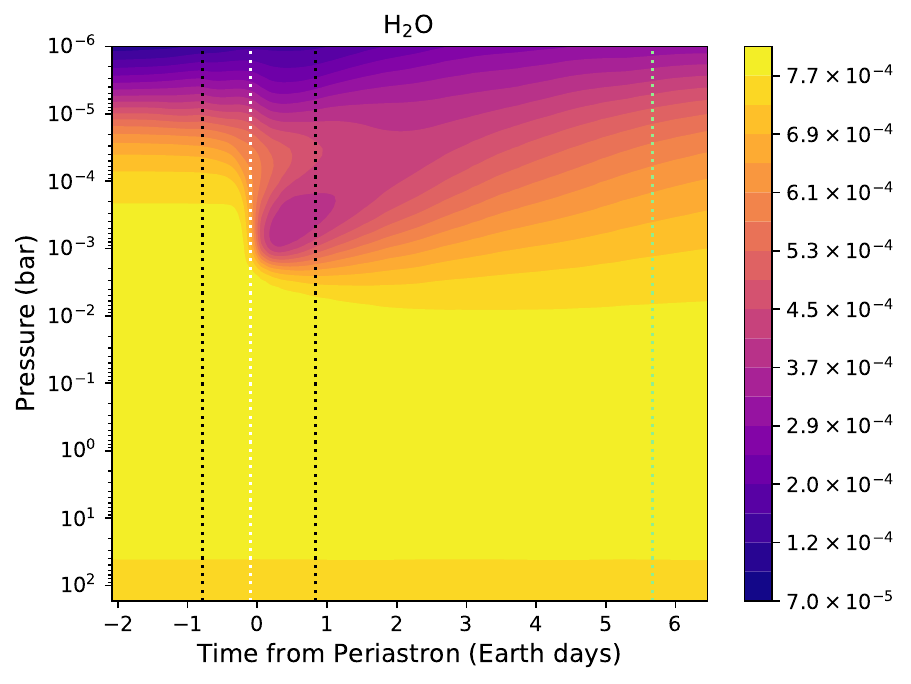}
    \includegraphics[width=\columnwidth]{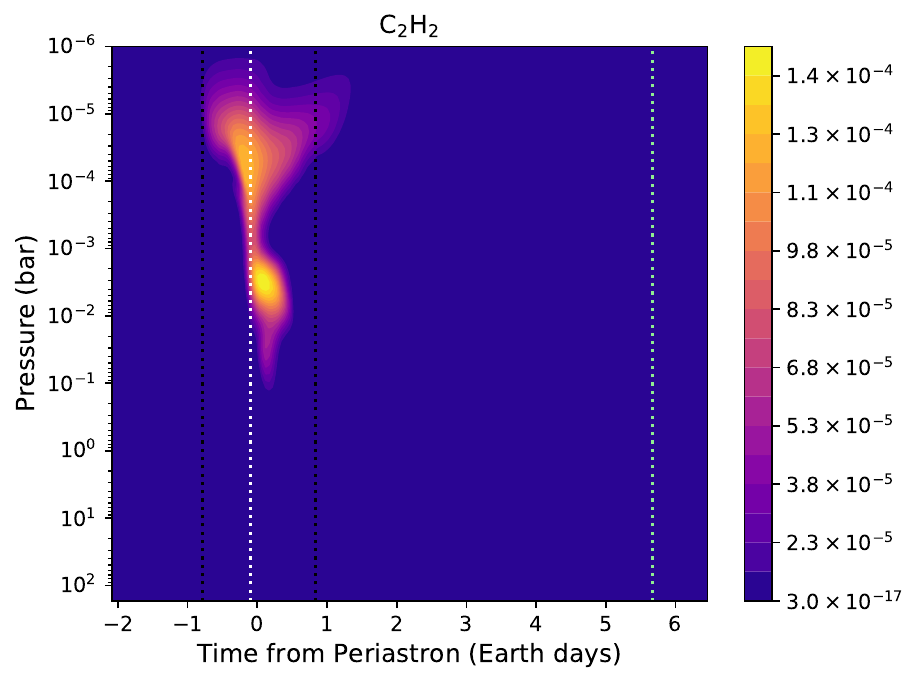}
    \includegraphics[width=\columnwidth]{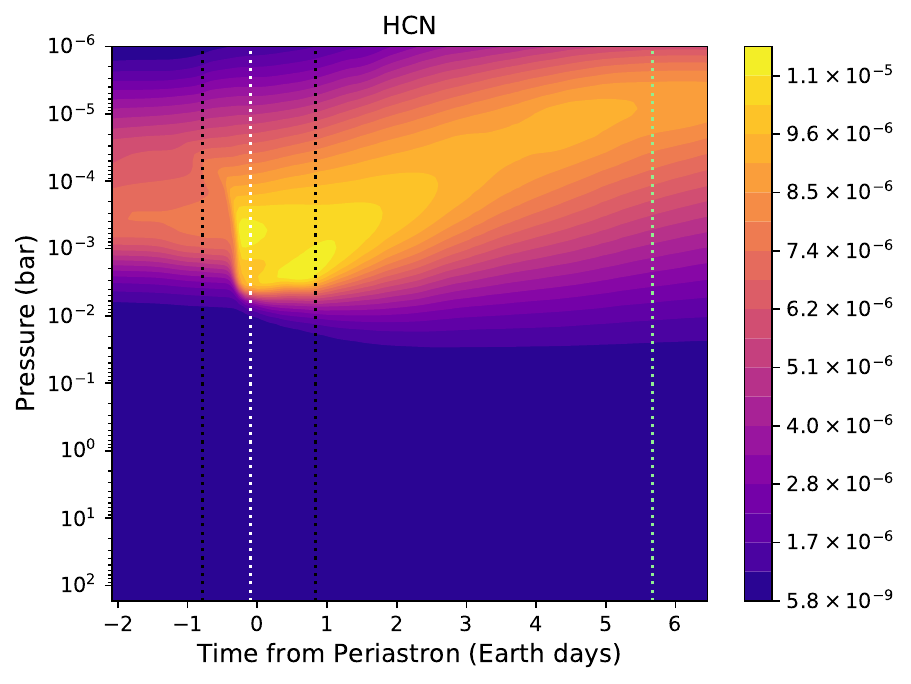}
      \caption{The dayside volume mixing ratios for several species of interest as a function of pressure and time (zero at periastron), showing the compositional response during the periastron passage. Note that the colors are shown on a linear scale. Following Figure \ref{fig:TPs}, the white dotted line indicates secondary eclipse while the black dotted lines enclose the synchronous-rotation period and the lime dotted line denotes the primary transit. }
         \label{fig:peri-contour} 
\end{figure*}

\subsubsection{Photochemistry-driven response near periastron}\label{sec:response_peri}

Photochemistry produces abundant radicals, which can drive atmospheric composition to respond much faster (than what pure thermal kinetics allows) in the upper atmosphere. We now examine how the composition varies in response to the sudden shift of incident stellar flux and temperature during the periastron passage. The compositional response is depicted by the mixing ratio profiles as a function of time in Figure \ref{fig:peri-contour}.

\begin{figure}
   \centering
   \includegraphics[width=\columnwidth]{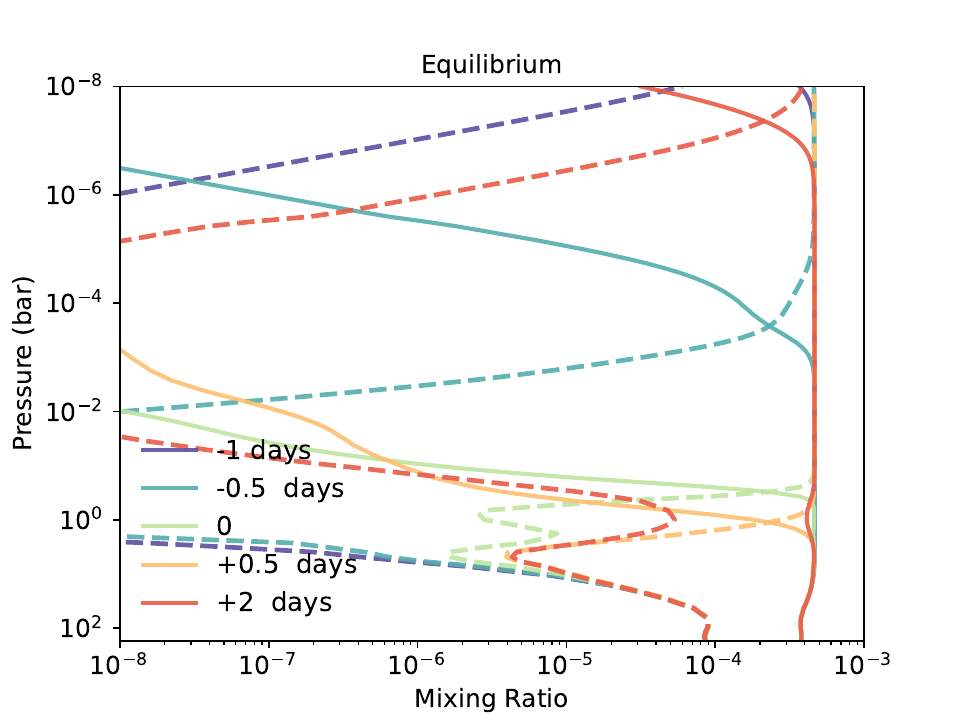}
   \includegraphics[width=\columnwidth]{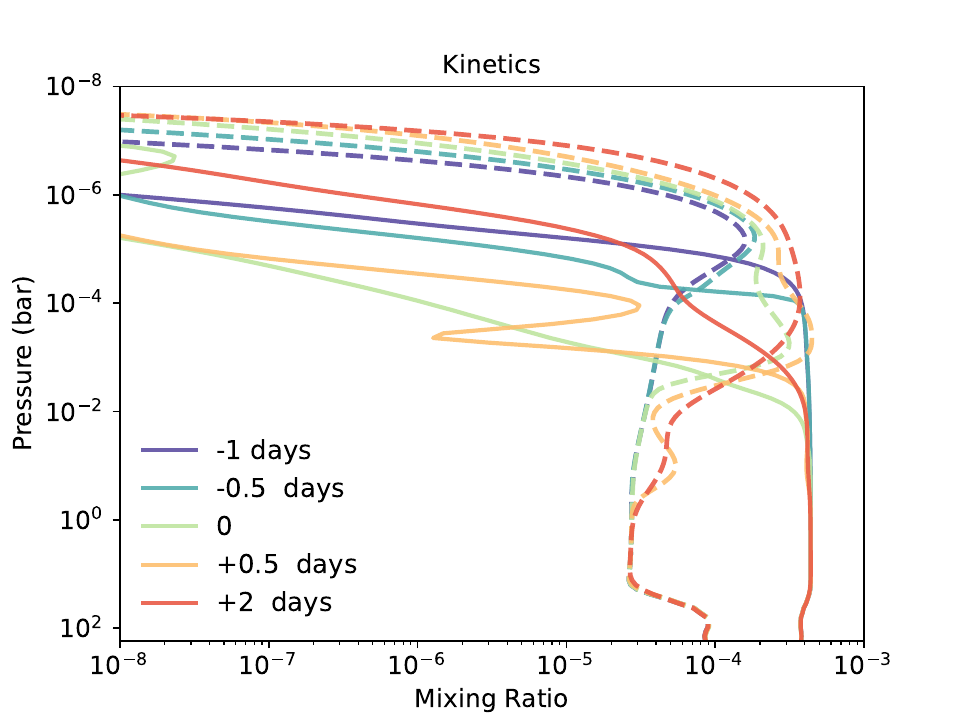}
      \caption{Snapshot of equilibrium (top) and disequilibrium (bottom) mixing ratio profiles of \ce{CH4} (solid) and CO (dashed) at different times near periastron in our nominal HD 80606 b model. The violet to orange colors denote -1, 0.5, 0, 0.5, and 2 days with respect to periastron.}
         \label{fig:CH4_CO_1Ds}
\end{figure}

The stratosphere between $\sim$ 1 and 10$^{-4}$ bar quickly heats up when the planet is approaching periastron. It becomes warm enough (about half day before and one day after periastron) such that CO would have dominated over \ce{CH4} above 1 bar in thermochemical equilibrium, as suggested by Figure \ref{fig:timescale} and confirmed in Figure \ref{fig:CH4_CO_1Ds}. However, the duration of this hot state is orders of magnitude shorter than the CO--\ce{CH4} chemical timescale (Figure \ref{fig:timescale}). We find photochemically produced radicals, such as H, OH, and S, to be key to promptly initiate the \ce{CH4}--CO conversion. After periastron, the produced CO mixing ratio can exceed 10$^{-4}$ for a few days before getting dissipated by vertical mixing. An intriguing feature is that the slope of CO abundance in time in Figure \ref{fig:peri-contour} essentially captures the lifetime of the transient CO.

\begin{figure}
   \centering
   \includegraphics[width=\columnwidth]{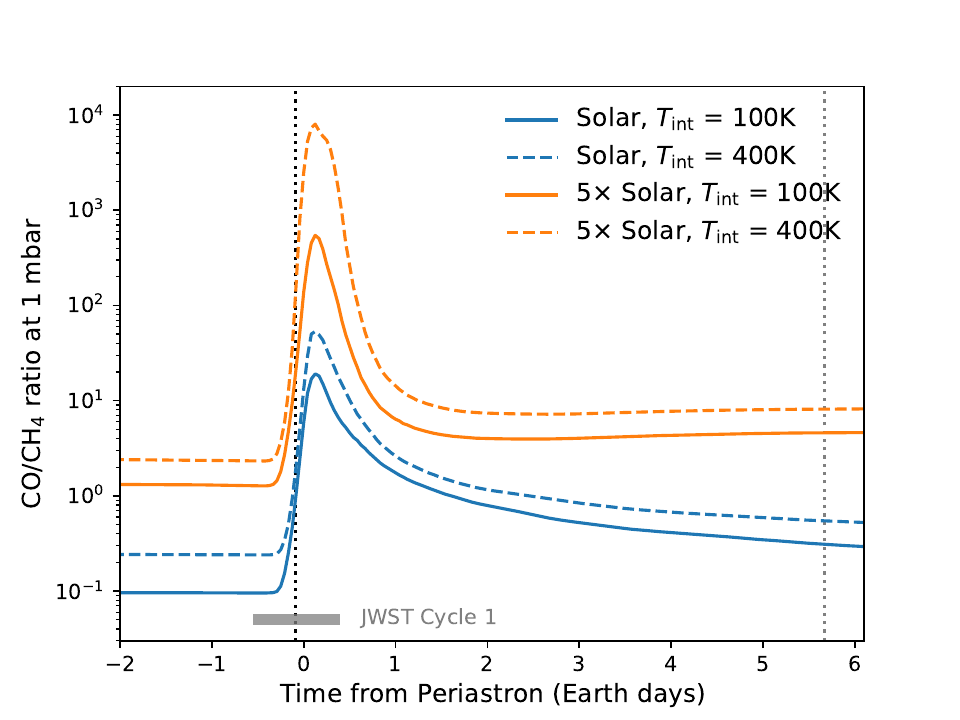}
      \caption{[CO]/[\ce{CH4}] ratios around the periastron passage for different assumptions of atmospheric metallicity and internal temperature in our HD 80606 b model. The black and grey dotted lines indicate the secondary eclipse and primary transit, respectively. The scheduled observation time with JWST Cycle 1 program is depicted by the bottom grey band.}
         \label{fig:CO_CH4_ratio}
\end{figure}

The variation of [CO]/[\ce{CH4}] ratio at 1 mbar is further illustrated in Figure \ref{fig:CO_CH4_ratio}. [CO]/[\ce{CH4}] starts to rise just before the eclipse and peaks about 8 hours after periastron, where [CO]/[\ce{CH4}] $\gtrsim$ 10 -- 10$^4$, depending on the atmospheric metallicity and internal temperature. The JWST Cycle 1 observations probing the atmosphere above $\sim$ 10 mbar, especially by NIRSpec where \ce{CH4} and CO have strong features within this wavelength range, will potentially provide the first real-time track of \ce{CH4}--CO conversion.


We also find that the variation of \ce{H2O} inversely follows CO, as oxygen is transferred from \ce{H2O} to CO in the net reaction \ce{CH4 + H2O -> CO + 3H2}. In addition, photochemical products, such as HCN and \ce{C2H2}, are promoted near periastron and have significantly higher concentrations than their equilibrium abundances. The production of HCN follows the shared dissociation levels of \ce{CH4} and \ce{NH3} \citep{Moses11,Tsai2021} and dissipates similarly to CO after periastron. \ce{C2H2} briefly reaches a high concentration of 10$^{-4}$ across the atmosphere (0.1 -- 10$^{-5}$ bar) around periastron. The transient peak of \ce{C2H2} is a combination of \ce{C2H2} being photochemically produced and thermodynamically favored at temperature $\gtrsim$ 1000 K around periastron . Conversely, \ce{CO2} momentarily falls back to the lower abundance in chemical equilibrium by about two orders of magnitude during this period. The peak of \ce{CO2} in the upper atmosphere is restored just before transit, following the recovery \ce{H2O} which produces OH radical turning CO into \ce{CO2}. Lastly, while most substantial variation of the main carbon, oxygen, and nitrogen species occur near periastron, the overview of the variation of their abundances along the full orbit can be found in Figure \ref{fig:full-contour}.





\begin{figure*}
   \centering
   \includegraphics[width=\columnwidth]{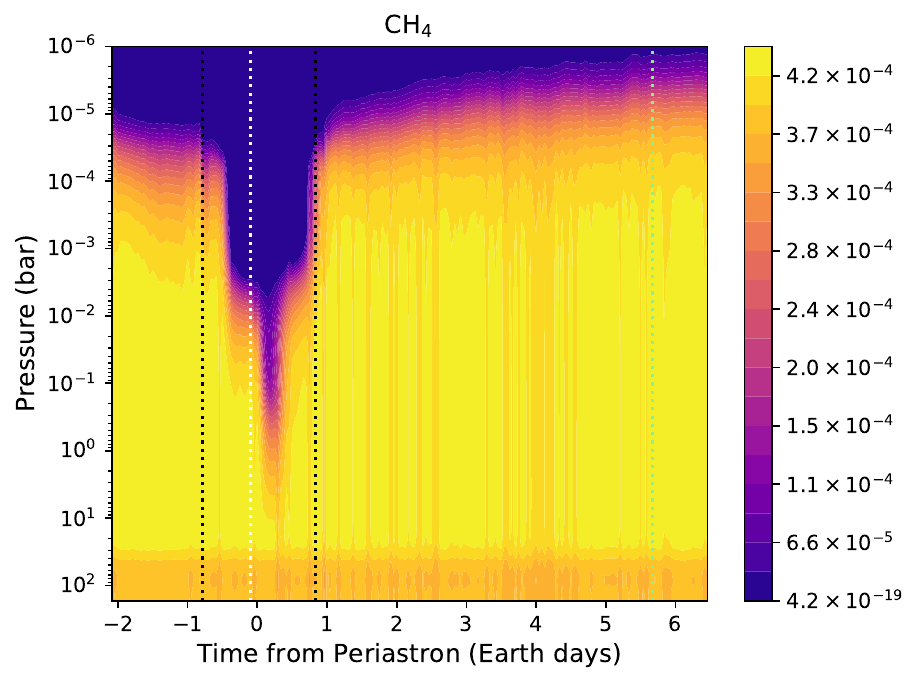}
   \includegraphics[width=\columnwidth]{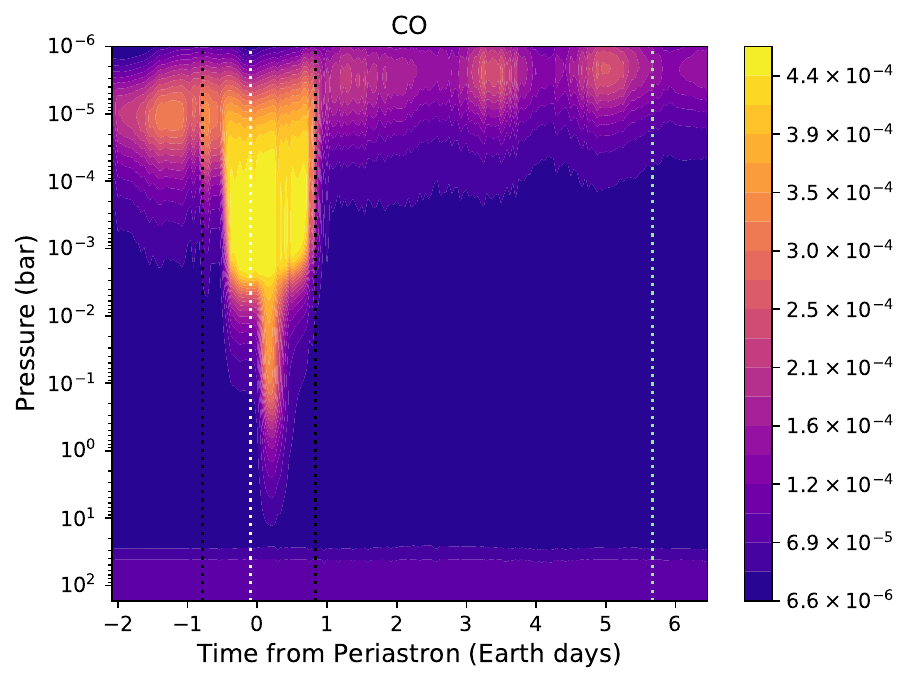}
   \includegraphics[width=\columnwidth]{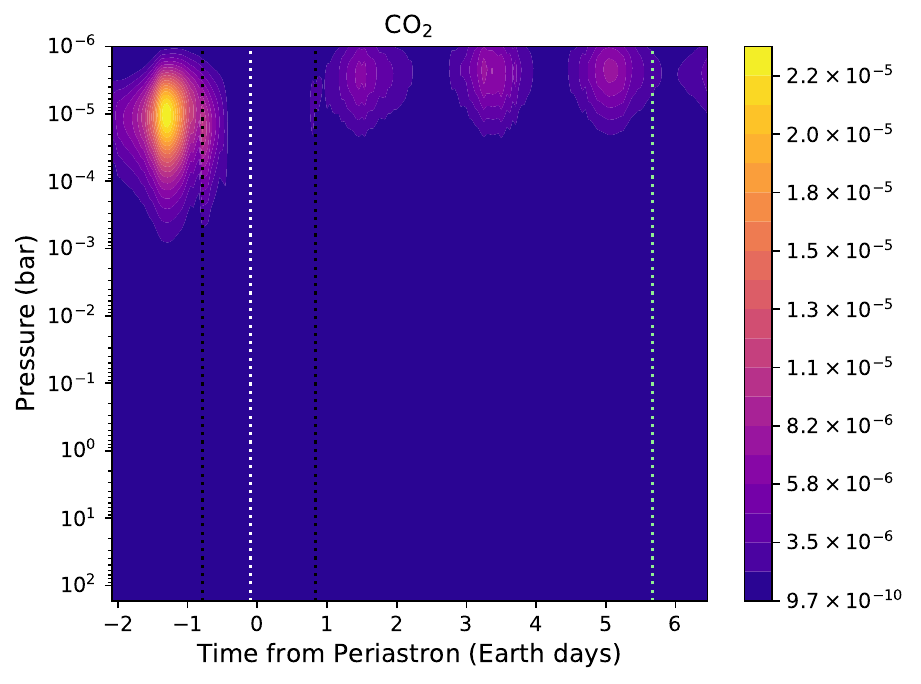}
   \includegraphics[width=\columnwidth]{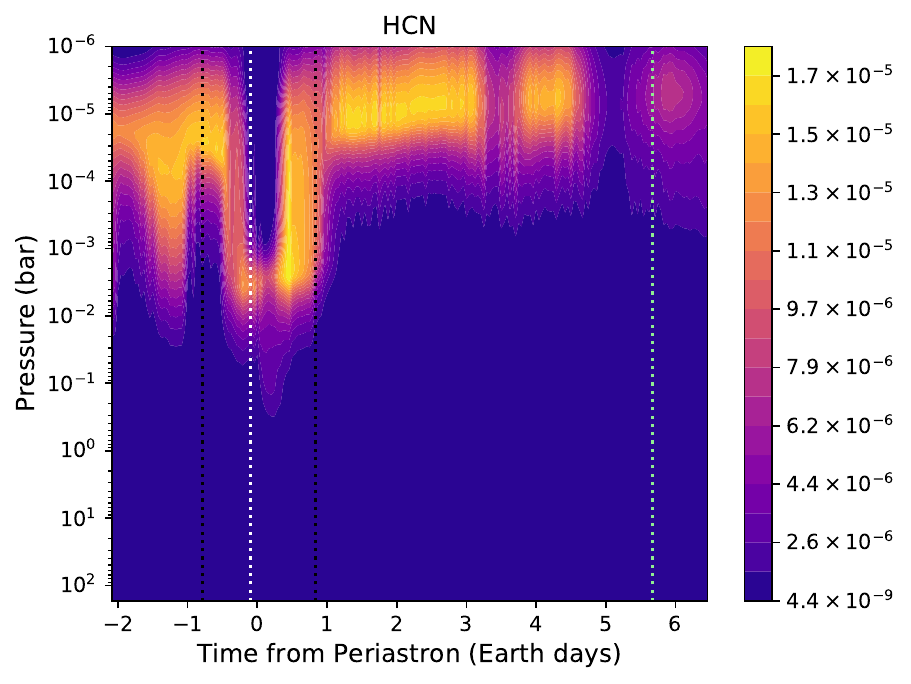}
   \caption{The same dayside volume mixing ratios for some species from Figure \ref{fig:peri-contour} except the planet in our photochemical model is fixed at each orbital position and allowed to reach steady state.}\label{fig:peri-contour-noOrbit}
\end{figure*}

\subsubsection{The effect of orbit-induced quenching}\label{sec:1D-orbit} 
To further understand how the eccentric orbit impacts the composition variations, we perform control experiments to isolate the effects of orbital motion. In this setup, we rerun the same photochemical model at each orbital position except keeping the planet fixed, allowing the composition to evolve until the steady state has reached. Figure \ref{fig:peri-contour-noOrbit} provides the mixing ratio profiles at the same orbital positions as Figure \ref{fig:peri-contour}.

Figure \ref{fig:peri-contour-noOrbit} shows that with enough time, \ce{CH4} can fully convert to CO above $\sim$ 10 mbar within the synchronous-rotation period in the nominal case (about 1 day before and after periastron). The depletion of \ce{CH4} subsequently restrains the production of HCN, making the variation of HCN completely differ from the continuous response in the nominal model with orbital motion. Outside of the synchronous-rotation period, \ce{CH4} remains the dominant carbon-bearing molecule. The elevated vertical mixing after periastron is somewhat more favorable for \ce{CH4} than CO and subsequently suppresses \ce{CO2} at steady state. This further confirms that in the nominal model with an eccentric orbit, the peak of \ce{CO2} around transit is a result of the leftover CO. 

\begin{figure*}
   \centering
   \includegraphics[width=0.9\columnwidth]{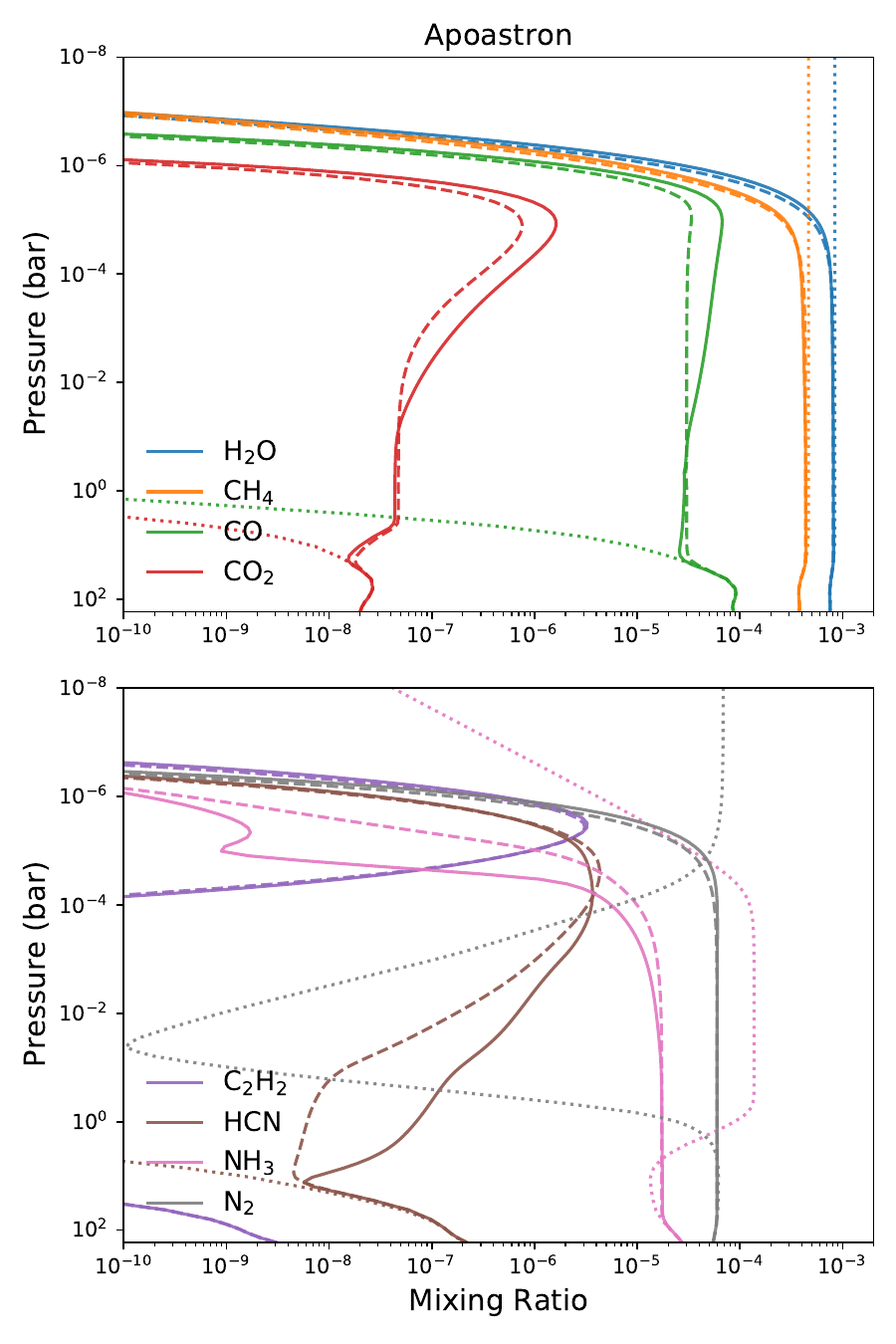}
   \includegraphics[width=0.9\columnwidth]{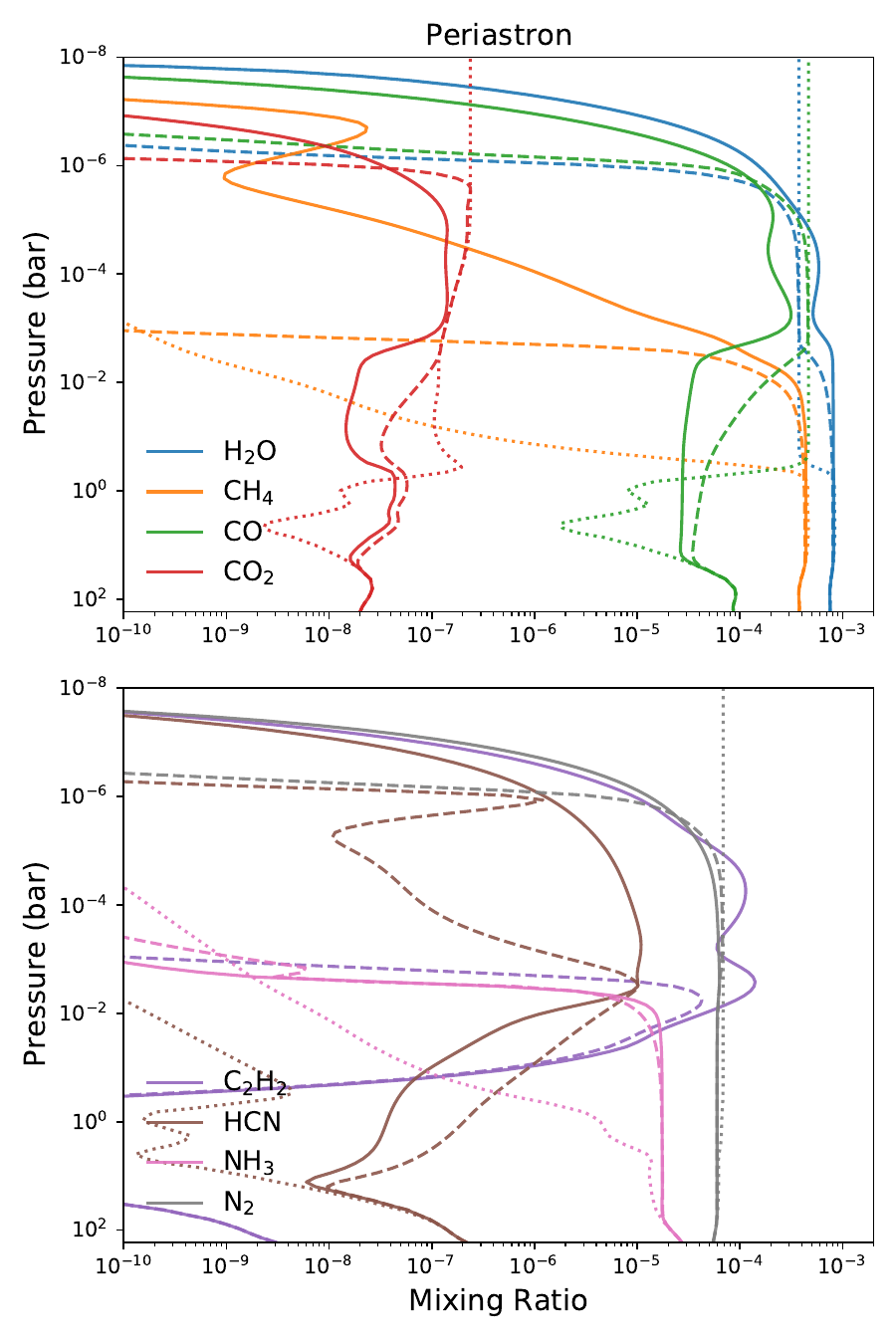}
      \caption{Composition profiles of HD 80606 b in our nominal model (solid) and the fixed-orbit model (dashed), at apoastron (left) and periastron (right). The dotted lines show the equilibrium abundances.}
      \label{fig:1D-orbit}
\end{figure*}

We highlight the role of orbit-induced variations in Figure \ref{fig:1D-orbit} by comparing the nominal model, fixed-orbit model, and chemical equilibrium for apoastron and periastron. While the difference between the equilibrium abundance and the fixed-orbit model informs us the effect of disequilibrium chemistry alone, the comparison between the nominal model and the fixed-orbit model allows us to tease out the orbital effects.

The close match between the nominal and fixed-orbit models in the left panel of Figure \ref{fig:1D-orbit} indicates that the orbit-induced variations has minimal effect at apoastron, since the irradiation and temperature vary little away from periastron. In this cold state, CO maintained a quenched mixing ratio around 10$^{-5}$, with a small amount of leftover from the hot state. It is predominantly vertical quenching, which is generally time independent (see the  discussion in \ref{sec:st_quench}), that controls the major species with little contribution from the orbital motion. On the other hand, the atmosphere experiences rapid heating and a surge in the stellar flux over a few days during the periastron passage. The eccentric orbit motion takes over to dictate the thermal and chemical variations in this hot state. This can be seen in the right panel of Figure \ref{fig:1D-orbit} that several main species undergo rapid change but do not have enough time reach steady state. CO production initiates around 1 mbar while the residual \ce{H2O} and \ce{CH4} from the cold state are partially destroyed. An important consequence is that under the intense UV flux, the excess \ce{CH4} makes \ce{C2H2} and HCN much more abundant above 10 mbar due to orbit-induced quenching. 


\begin{figure*}
   \centering
    \includegraphics[width=\columnwidth]{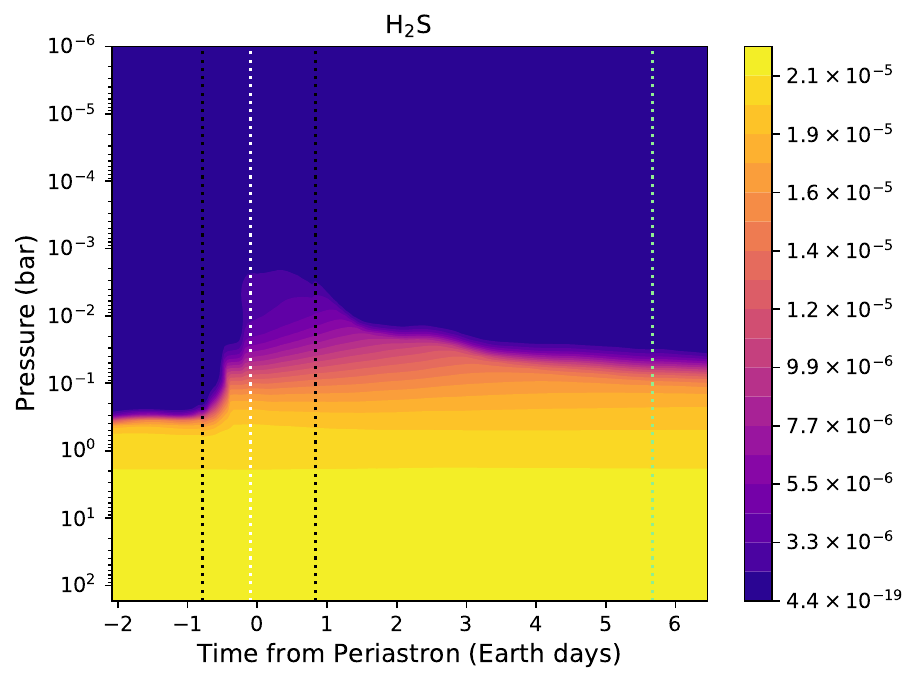}
    \includegraphics[width=\columnwidth]{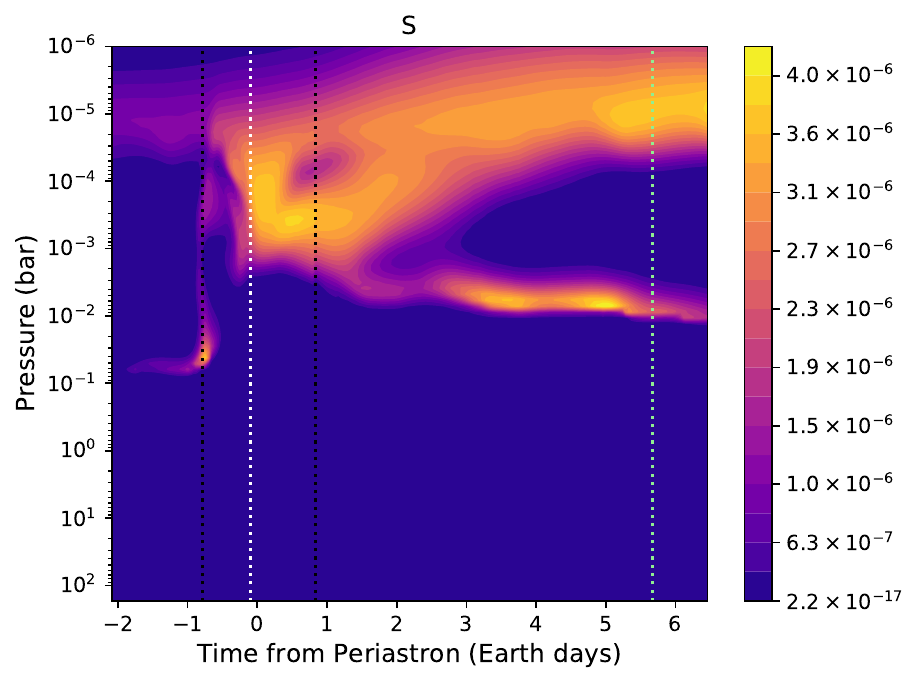}
    \includegraphics[width=\columnwidth]{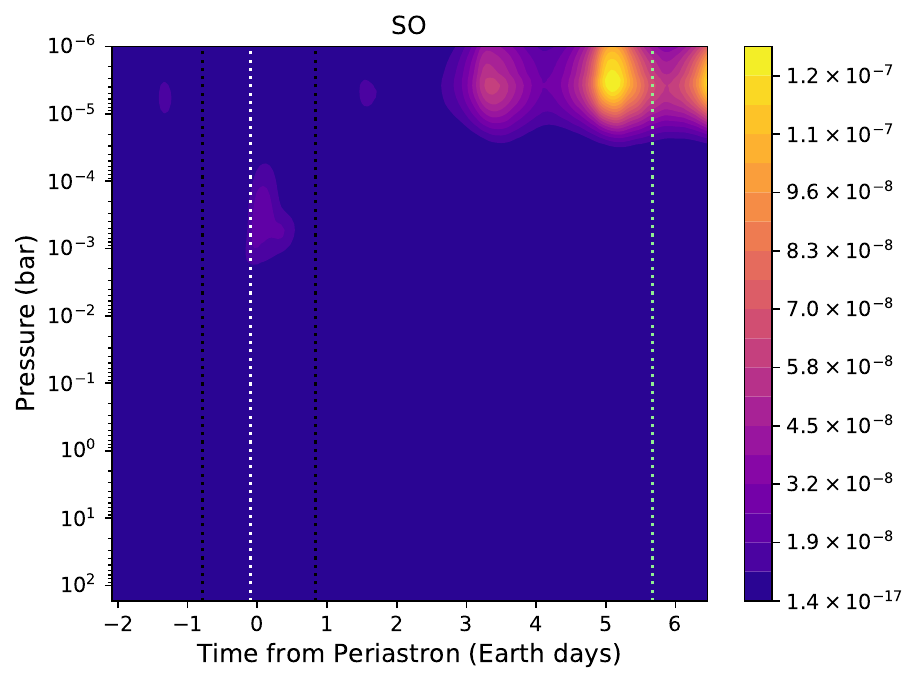}
    \includegraphics[width=\columnwidth]{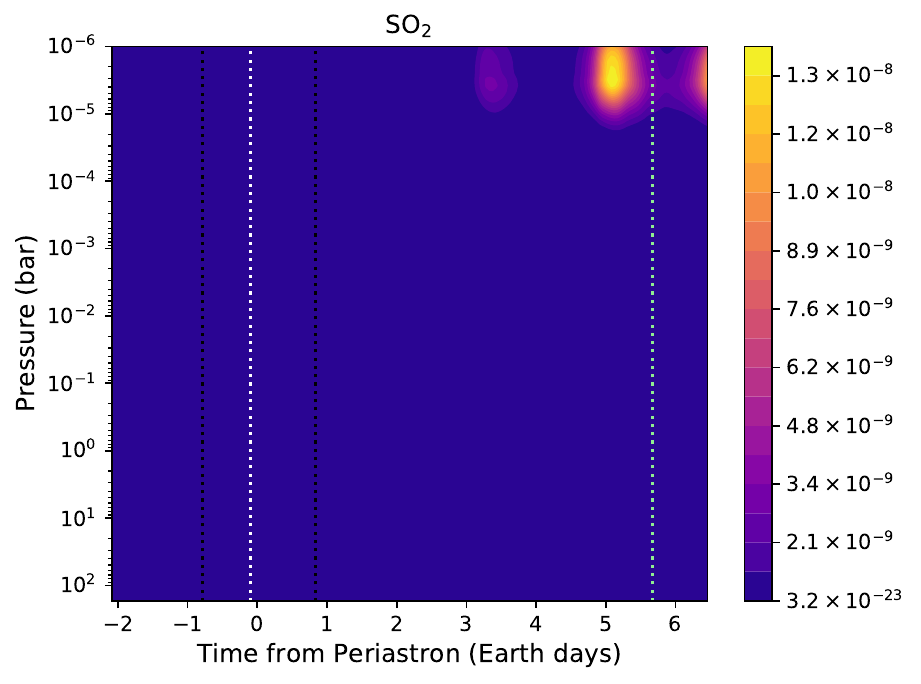}
    \includegraphics[width=\columnwidth]{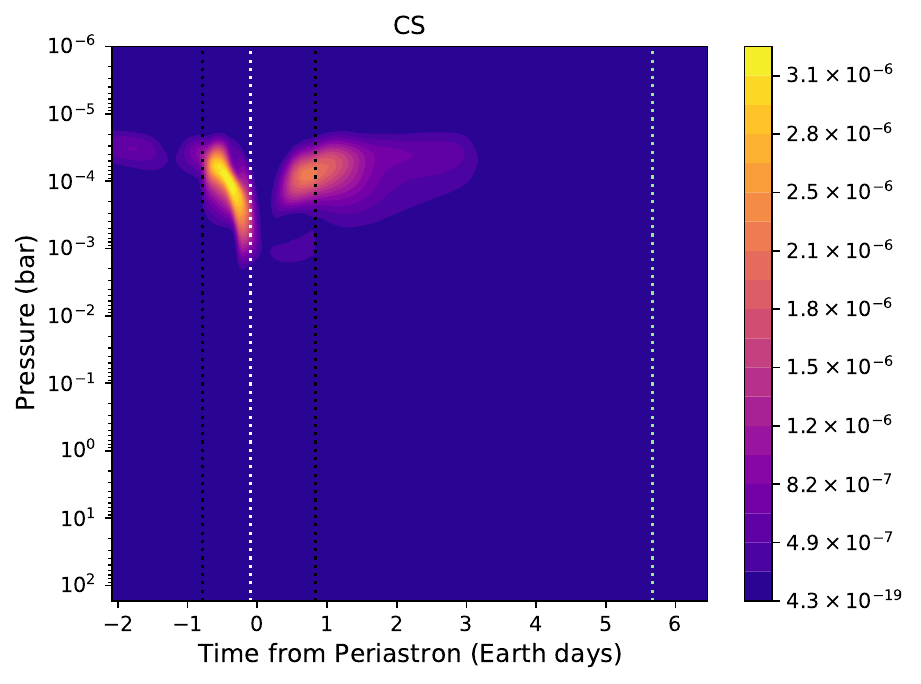}
    \includegraphics[width=\columnwidth]{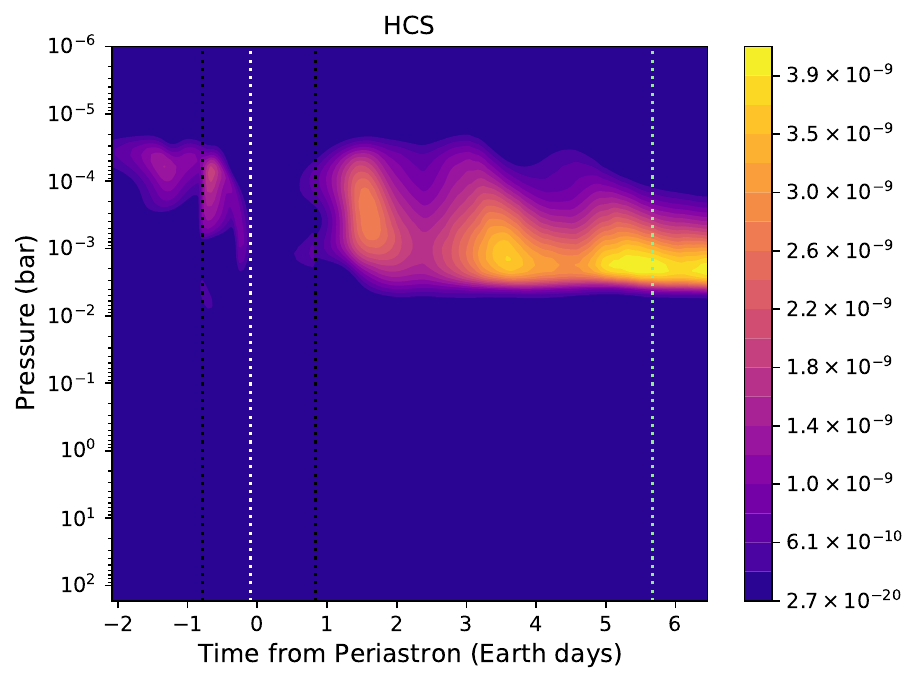}
    \caption{Same as Figure \ref{fig:peri-contour} but for sulfur species.} 
\label{fig:peri-contour-S}
\end{figure*}
\subsubsection{The short-term and long-term response of sulfur species}
\begin{figure*}
   \centering
    \includegraphics[width=\columnwidth]{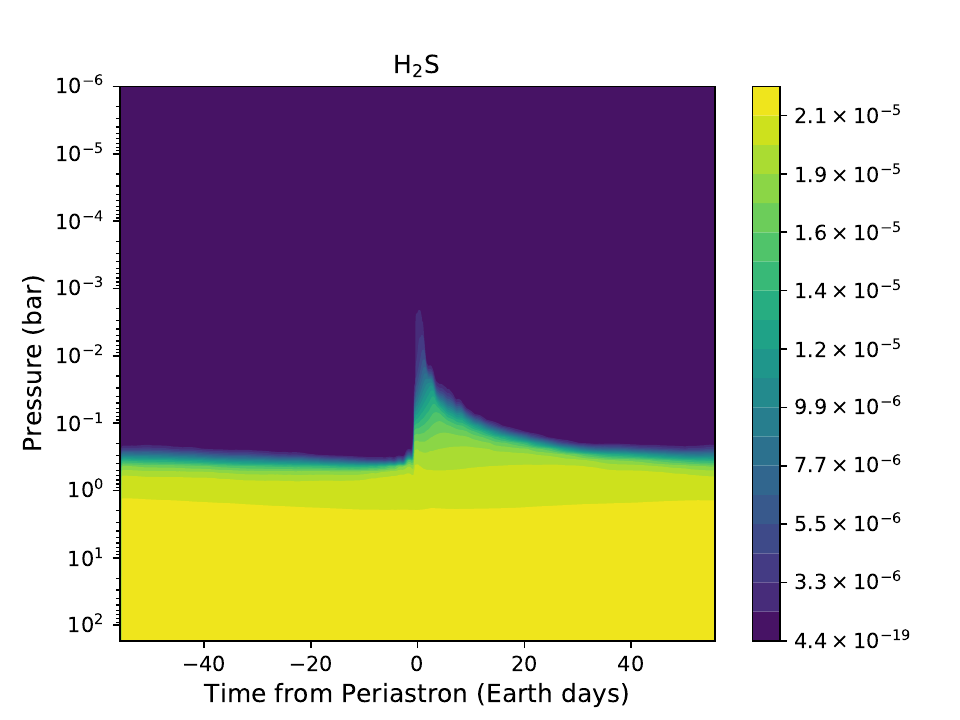}
    \includegraphics[width=\columnwidth]{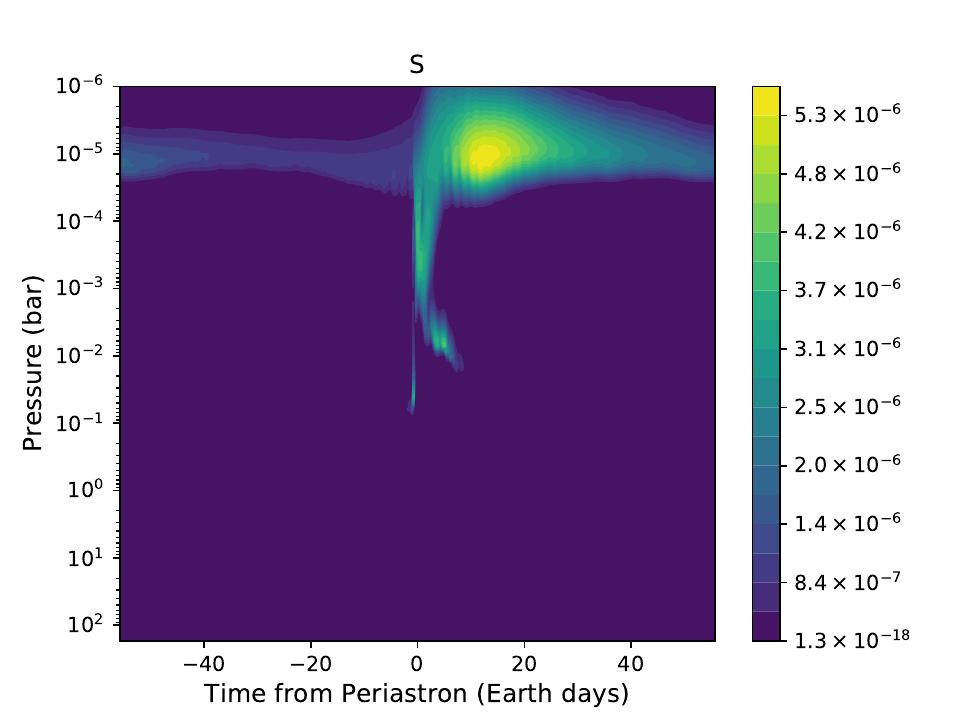}
    \includegraphics[width=\columnwidth]{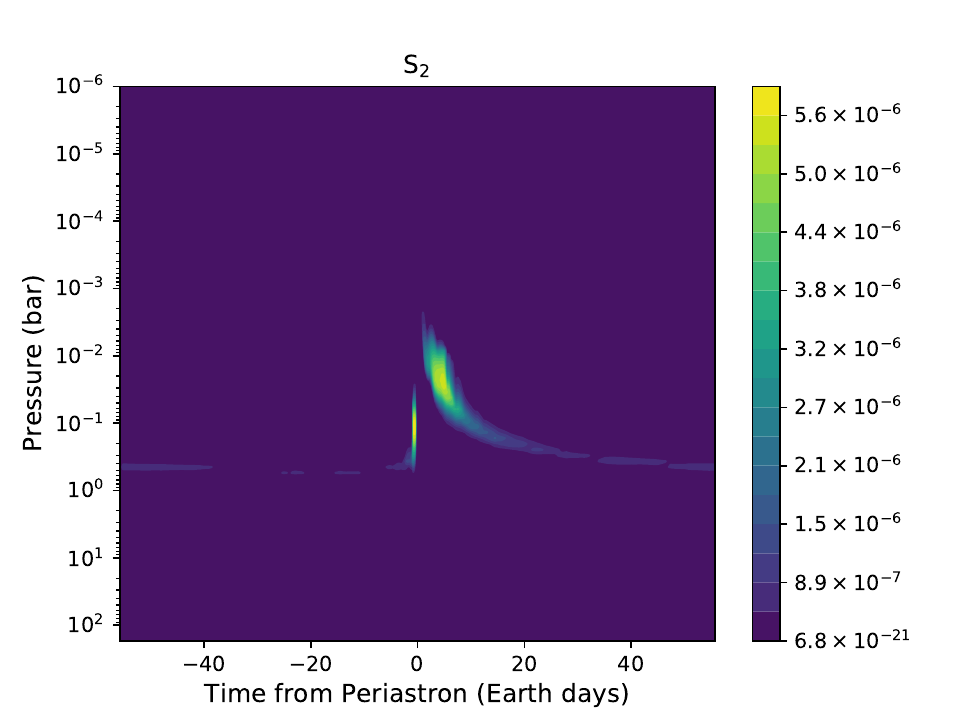}
    \includegraphics[width=\columnwidth]{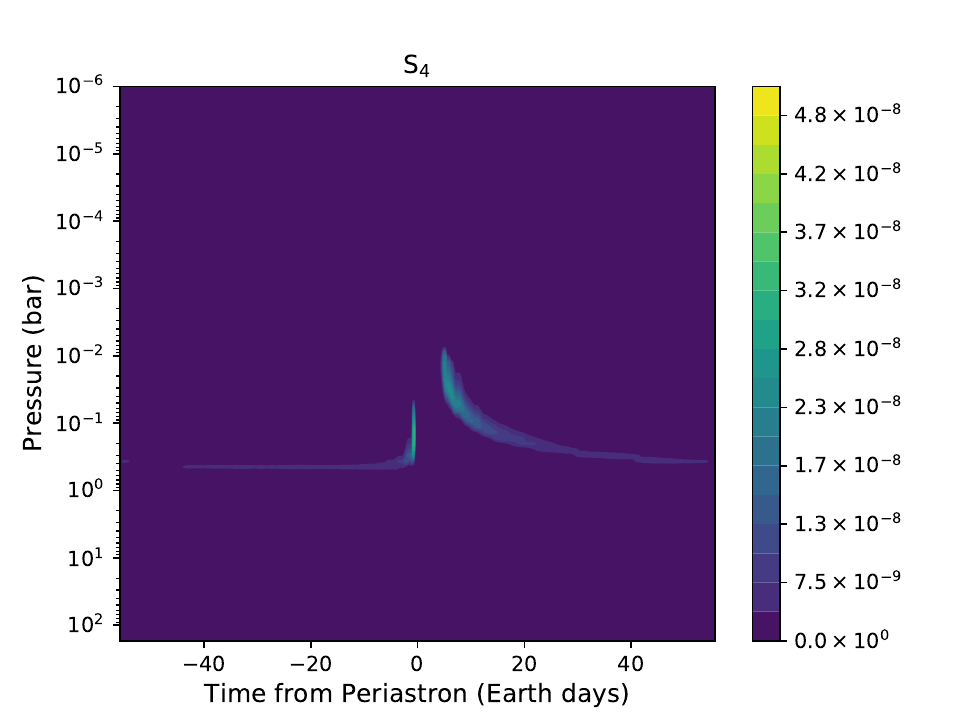}
    \includegraphics[width=\columnwidth]{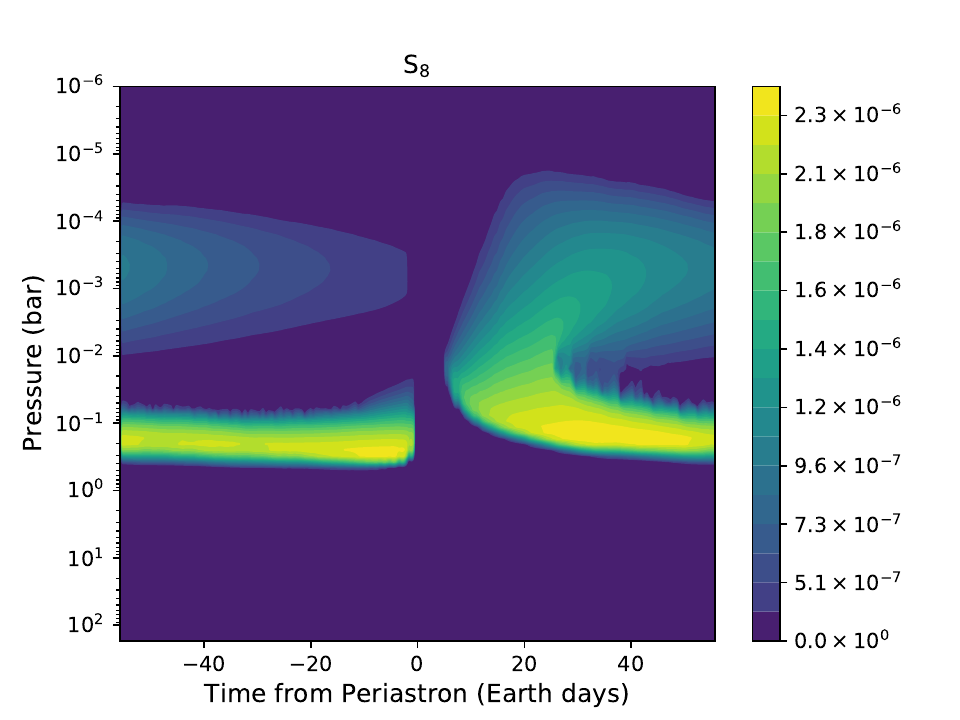}
    \includegraphics[width=\columnwidth]{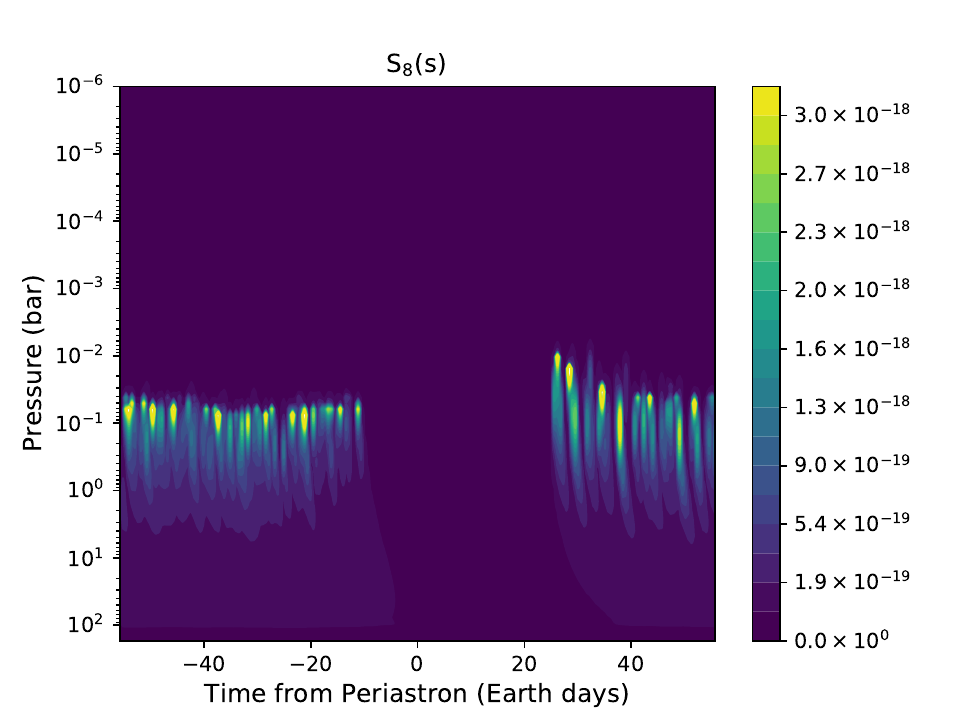}
    \caption{The full-orbit evolution of the dayside composition of several sulfur species that form allotropes and hazes. \ce{S8} particle mixing ratio is shown in n$_{\ce{S8}}$/n$_{\textrm{gas}}$, where n$_{\textrm{gas}}$ is the number density (cm$^{-3}$) of the total gas. Note the elemental sulfur cycle consisted of S--\ce{S2}--\ce{S4}--\ce{S8} sequence that spanned the orbit.}
\label{fig:full-contour-S}
\end{figure*}

We have discussed the behavior of carbon, oxygen, and nitrogen species so far. Next, we will examine the response of sulfur species. Figure \ref{fig:peri-contour-S} presents the dayside composition variation of several important sulfur species near periastron while that of the full orbit is shown in Figure \ref{fig:full-contour-S}. 

\ce{H2S} is the thermodynamically stable sulfur bearing molecule in a \ce{H2} atmosphere. Under shortwave irradiation where H atoms are produced in the upper atmosphere, \ce{H2S} is attacked by H and branches to S or SH \citep{Zahnle2016}. The reaction that recycle sulfur back to \ce{H2S},  
\begin{equation}
\ce{SH + H2 -> H2S + H},
\label{re:H2S}
\end{equation}
has a high energy barrier and strong temperature dependence. For instance, the rate constant of Reaction (\ref{re:H2S}) at about 0.1 bar increases from 10$^{-20}$ cm$^{-3}$ s$^{-1}$ to 10$^{-13}$ cm$^{-3}$ s$^{-1}$ as temperature climbs from 400 K to 1300 K near apoastron. Therefore, \ce{H2S} can remain stable at higher altitudes with rising temperatures. The stable level of \ce{H2S} tracks the thermal structure around periastron (Figure \ref{fig:TPs}) and remains elevated for a few days after periastron. The elevated abundance of \ce{H2S} is important in bringing it to the H-rich photolysis region and initiating the production of elemental S \citep{Tsai2021}, which sets up both short-term and long-term evolution along the orbit. 

For the short-term response during periastron passage, S is oxidized to SO and \ce{SO2} in the upper atmosphere where OH is abundant, reaching a peak just before transit. Unlike other main species that respond primarily to the rapid irradiation near periastron and exhibit little variation for the rest of the orbit, sulfur also undergoes long-term cycles across the orbit, where it grows into elemental allotropes in the cold and reducing conditions. The evolution of sulfur species for a full orbit are illustrated in Figure \ref{fig:full-contour-S}. After periastron passage, the temperature gradually cools down and the lower UV flux allows the upper atmosphere to resume a more reducing condition. S starts to from allotropes in sequence: Once abundant atomic S is liberated from \ce{H2S} by photochemistry, \ce{S2} can form via either \ce{S + SH -> S2 + H} or \ce{S + H2S -> S2 + H2}. \ce{S2} then continues with third-body assisted self reactions to form bigger allotropes (S$_x$). These growth steps following \ce{S -> S2, S3 -> S4 -> S8} are captured in Figure \ref{fig:full-contour-S}. We find \ce{S8} to be the main sulfur-bearing species in the gas phase above 0.1 bar outside the periastron passage. \ce{S8} becomes saturated and condenses to form sulfur hazes between 1 and 0.01 bar with temperatures lower than about 360 K \citep{Gao2017,Tsai2021}. This long-term sulfur cycle is remotely analogous to the photochemical evolution after Shoemaker-Levy 9 impacts and shocks on Jupiter \citep{Moses1995,Moses1996}.



\begin{figure}
   \centering
   \includegraphics[width=\columnwidth]{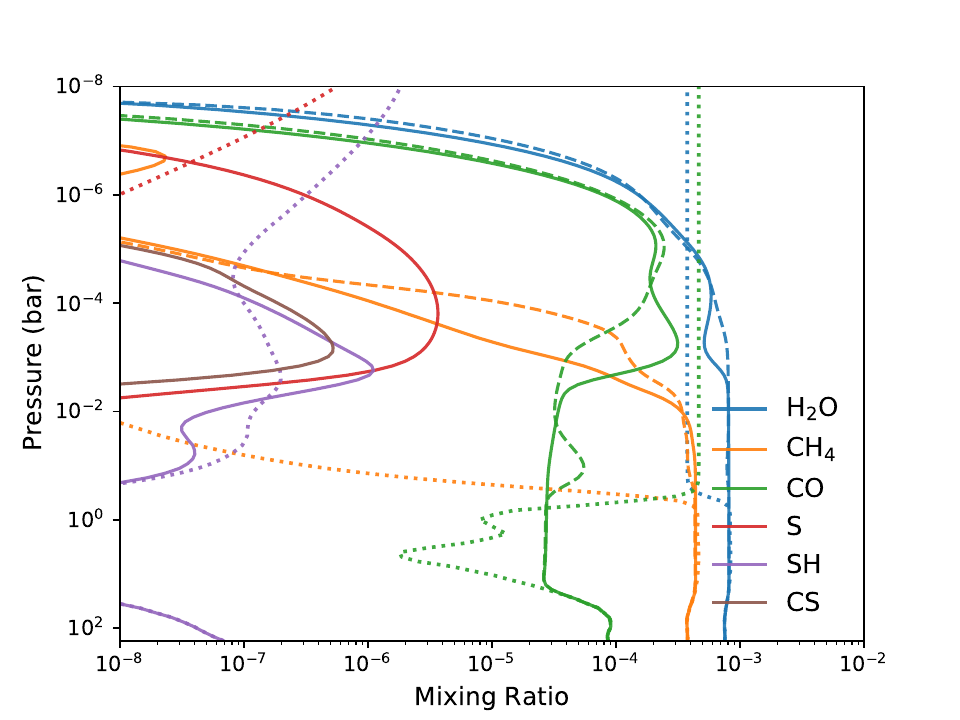}
      \caption{The composition profiles of HD 80606 b six hours after periastron in our nominal model (solid) and the same model but without sulfur (dashed), while the dotted lines indicate the equilibrium abundances.}
      \label{fig:peri06-noS}
\end{figure}

\subsubsection{The impact of sulfur on other species}
\cite{Tsai2021} identify that sulfur can impact other species in a nonlinear way, including accelerating \ce{CH4}--CO conversion with catalytic paths. Since \ce{CH4} can only partially convert to CO near periastron, HD 80606 b provides an ideal setting to test the effects of sulfur on carbon conversion. To this end, we performed our nominal model except without including sulfur chemistry. Figure \ref{fig:peri06-noS} compares the abundances from the models with and without sulfur six hours after periastron, when the \ce{CH4}--CO conversion reaches maximum. Sulfur participation significantly enhances CO production between 10$^{-2}$ bar and 10$^{-4}$ bar, producing the peak of CO seen in Figure \ref{fig:peri-contour}. 

We performed pathway analysis to look into the role of sulfur participation. Without sulfur, the main pathway for \ce{CH4}--CO conversion is 
\begin{eqnarray}
\begin{aligned}
\ce{CH4 + H &-> CH3 + H2}\\
\ce{CH3 + H &-> CH2 + H2}\\
\ce{CH2 + H &-> CH + H2}\\
\ce{CH + H2O &-> H2CO + H}\\
\ce{H2CO + H  &-> HCO + H2}\\
\ce{HCO &->[M] H + CO}\\
2(\ce{H2O &->[h\nu] OH + H})\\
2(\ce{OH + H2  &-> OH + H})\\
\hline 
\noalign{\vglue 3pt}
\mbox{net} : \ce{CH4 + H2O &-> CO + 3H2}. 
\end{aligned}
\label{re:CH4-path}
\end{eqnarray}
where the reaction involving water to form \ce{H2CO} is the rate-limiting step. In the presence of sulfur, \ce{CH4} goes through a catalytic pathway, similar to that identified on HD 189733 b \citep{Tsai2021}:
\begin{eqnarray}
\begin{aligned}
\ce{CH4 + H &-> CH3 + H2}\\
\ce{CH3 + H &-> CH2 + H2}\\
\ce{CH2 + S &-> HCS + H}\\
\ce{HCS + H &-> CS + H2}\\
\ce{CS + OH &-> OCS + H}\\
\ce{OCS + H &-> CO + SH}\\
\ce{SH + H &-> S + H2}\\
2(\ce{H2O &->[h\nu] OH + H})\\
\ce{OH + H2  &-> H2O + H}\\
\hline 
\noalign{\vglue 3pt}
\mbox{net} : \ce{CH4 + H2O &-> CO + 3H2}.
\end{aligned}
\label{re:CH4-S-path}
\end{eqnarray}
where the rate-limiting step in (\ref{re:CH4-S-path}), \ce{CH2 + S -> HCS + H}, is about 3 orders of magnitude faster than the reaction producing \ce{H2CO} in (\ref{re:CH4-path}) at 1 mbar. This reaction and pathway (\ref{re:CH4-S-path}) rely on abundant S to be efficient, thus the peak of sulfur-enhanced conversion of \ce{CH4}--CO follows S as shown in Figure \ref{fig:peri06-noS}. 


\subsection{Sensitivity to atmospheric metallicity and internal heat}
We have seen that \ce{CH4}, instead of CO, remains the dominant carbon-bearing molecule for the majority of the orbit except for a few days during the periastron passage. Since \ce{CH4} is less favored at high metallicity and high temperature, we now address whether the general behavior holds when we increase atmospheric metallicity and internal heat. We also test the sensitivity to vertical mixing.

\subsubsection{Sensitivity to internal heat} 
The higher $T_{\textrm{int}}$ sets up a greater lapse rate in the deep layers below 10 bar, but the thermal and dynamical structure above 10 bar level remains insensitive to this increase in internal heat (Figure \ref{fig:TPs}). We are aware that the GCM might take a much longer time to converge in the deep region \citep{Carone2020,Mendonca2020,Wang2020}. Nevertheless, it is unlikely that the temperature and wind above 10 bar would be largely affected by the deep evolution. \ce{CH4} remains the dominant carbon-bearing molecule despite the higher quenched CO abundance. Overall, our $T_{\textrm{int}}$ = 400 K models show qualitatively similar thermal and chemical variations to the same models but with $T_{\textrm{int}}$ = 100 K, as seen in Figure \ref{fig:CO_CH4_ratio} (solid lines and dashed lines).

\subsubsection{Sensitivity to metallicity} 
The temperature in the photosphere rises when metallicity increased (Figure \ref{fig:TPs}), owing to the increased opacities \citep{Drummond2018}, mainly contributed by water. The combined thermal and chemical effects make the relative \ce{CH4} and CO ratio more sensitive to the change of metallicity than raising the internal heat alone.

Figure \ref{fig:1D-apo-peri-metallicity} compares the composition distributions with solar and five times solar metallicity at apoastron and periastron. When $T_{\textrm{int}}$ remained the same (upper panels), CO and \ce{CH4} have very close quenched abundances when metallicity increased by 5 times. During the periastron passage, CO takes over \ce{CH4} as the main carbon molecule above about 0.1 bar. When $T_{\textrm{int}}$ is raised to 400 K, the general trend remains the same but now the quenched CO abundance exceeds that of \ce{CH4}. An intriguing feature is that in both of our 5 times solar metallicity models, a second \ce{CH4}--CO conversion peak occurred deeper at about 0.1 bar due to stronger temperature inversion. As indicated in Figure \ref{fig:CO_CH4_ratio}, the maximum CO/\ce{CH4} ratios with five times solar metallicity are 1--2 orders of magnitude higher than their solar-metallicity counterparts. Overall, the parameters we explored suggest that both supersolar metallicity and high $T_{\textrm{int}}$ are required to have CO as the predominant carbon molecule over \ce{CH4} throughout the orbit. 





\subsubsection{Sensitivity to $K_{\textrm{zz}}$}\label{sec:Kzz} 
Regarding the parameterizing uncertainty of vertical mixing, we performed sensitivity tests with varying eddy diffusion. Figure \ref{fig:1D-Kzz} compares the models with the nominal eddy diffusion coefficient derived from the GCM's vertical wind to those scaled by 3 and 1/3. We find the quench levels remain not too sensitive to shifting $K_{\textrm{zz}}$. However, eddy diffusion can affect composition distributions in two major ways. First, vertical mixing controls the level where molecular diffusion becomes dominated and the species started to stratify following their own scale heights, i.e. homopause \citep{Leovy1982,Sinclair2020}. This stratification effect is seen in the decay of the water profiles in Figure \ref{fig:1D-Kzz}. Second, eddy diffusion controls the photochemical lifetime of the species in the stratosphere produced during the periastron passage. Stronger vertical mixing transports and dissipates these species more efficiently, resulting in the $K_{\textrm{zz}}$ dependence of the surplus CO at apoastron in Figure \ref{fig:1D-Kzz}.


\begin{figure}
   \centering
   \includegraphics[width=\columnwidth]{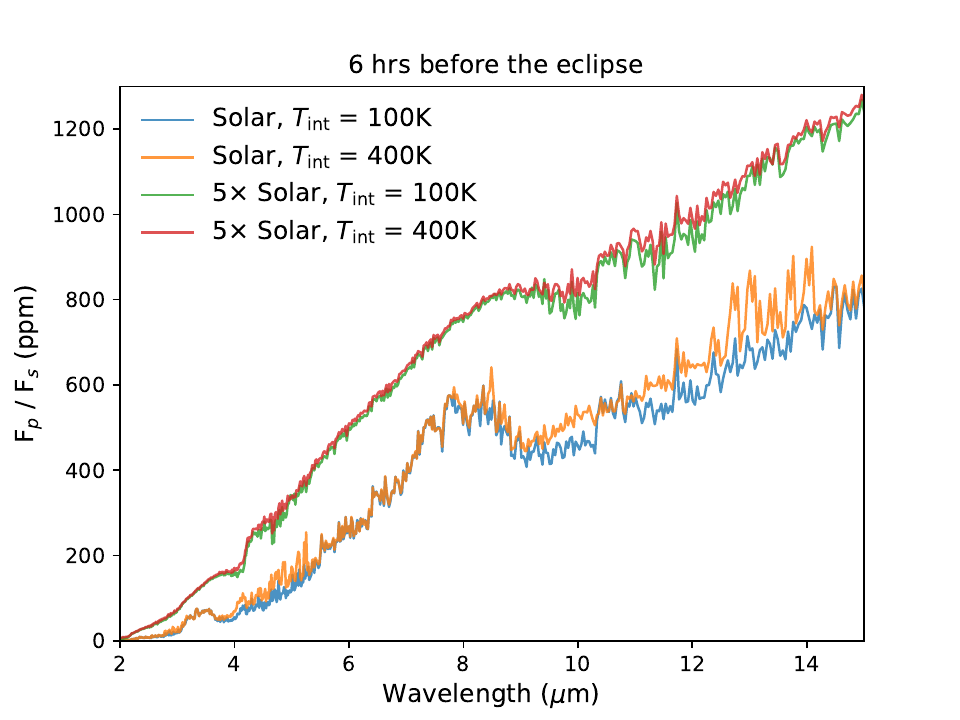}
   \includegraphics[width=\columnwidth]{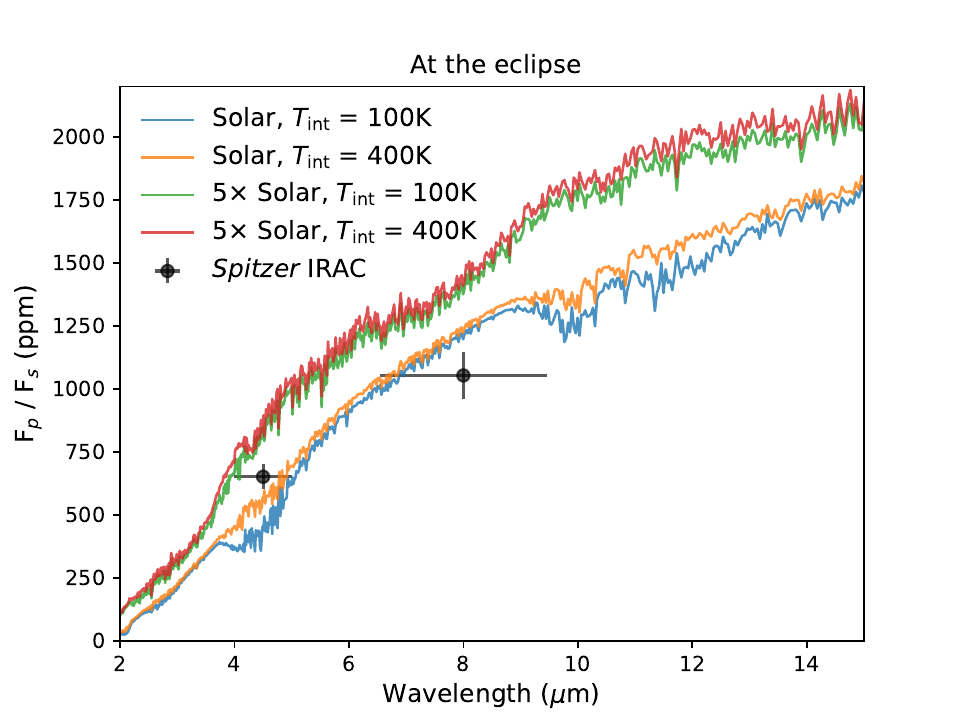}
   \includegraphics[width=\columnwidth]{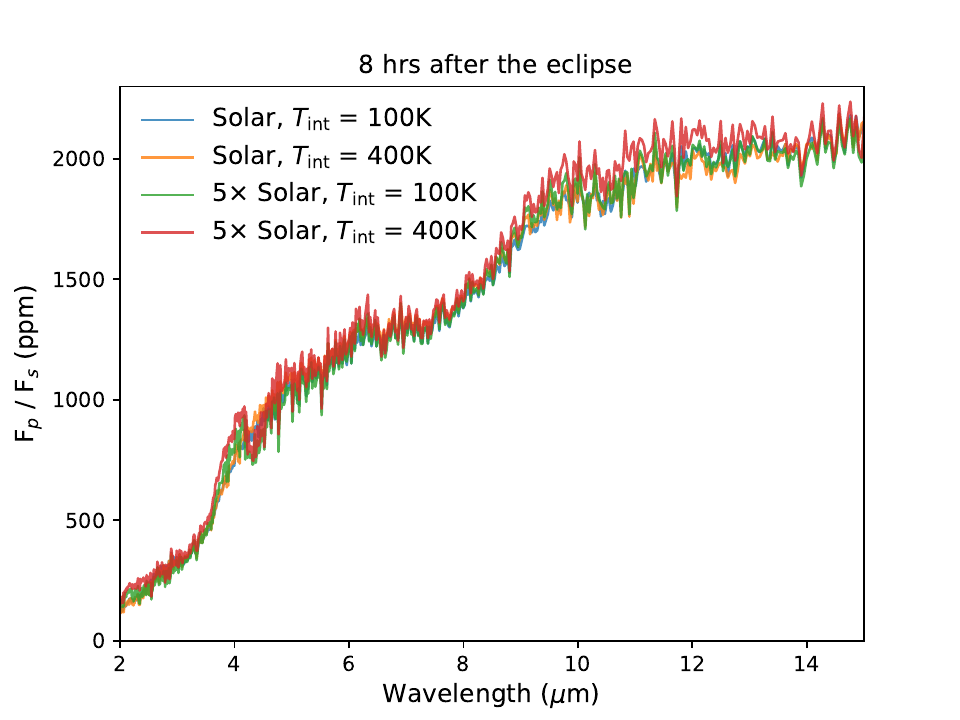}
      \caption{Emission spectra of HD 80606 b six hours before (upper), eight hours after (lower), and at the eclipse (middle), for models with different metallicities and internal heat. The data points in the middle panel are the eclipse depths observed by \textit{Spitzer} IRAC with Ch2 and Ch4 \citep{deWit2016}.}
      \label{fig:emission_4models}
\end{figure}

\subsection{Synthetic Spectra}\label{sec:spectra} 
In this section, we present the synthetic spectra generated by our 1D dayside average model. We do not take into account the change of Earth-facing hemisphere and the contribution from the nightside but focus on the shape and features of the spectra. Figure \ref{fig:emission_4models} compares the emission spectra of HD 80606 b six hours before, eight hours after, and at the secondary eclipse. We find that the emission flux is sensitive to the temperature increase with higher metallicity a few hours before and at the eclipse. During this period, the eclipse and pre-eclipse emission spectra are ideal for constraining the atmospheric metallicity. After the eclipse, the temperatures from models with different metallicities start to converge as the strong temperature inversion dominates the flux, leading to similar emission spectra. However, while the {\it Spitzer} measurement at 4.5 micron favors a slightly enhanced metallicity between solar and 5 $\times$ solar values, none of our models are able to fit the {\it Spitzer} IRAC Ch1 and Ch2 data simultaneously.



\begin{figure*}
   \centering
   \includegraphics[width=2\columnwidth]{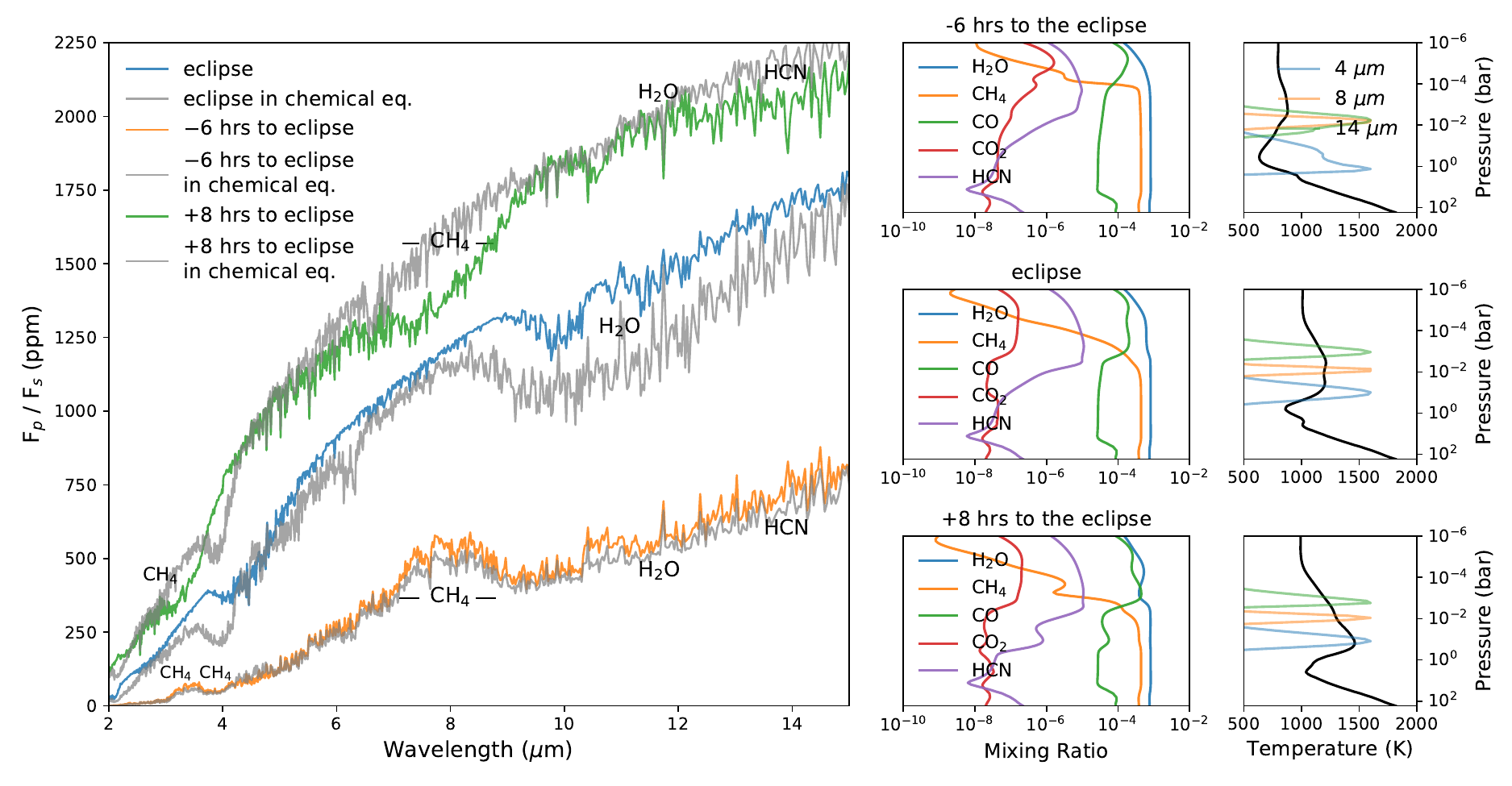}
      \caption{Emission spectra (left) of HD 80606 b six hours before the eclipse, at the eclipse, and eight hours after the eclipse, where the grey spectra are with equilibrium composition at each time. The right panels show the corresponding snapshots of the mixing ratios, temperature profiles, and contribution functions at 4, 8, and 14 micron. An animation of the full time-lapse spectra and atmosphere profiles is provided in the supplementary material.}
      \label{fig:emission_time}
\end{figure*}

Figure \ref{fig:emission_time} illustrates how the emission spectra of our nominal model evolved before and after the eclipse. Since the chemical conversion does not occur instantaneously, significant differences in the \ce{CH4} and \ce{H2O} band exhibit when assuming chemical equilibrium. We find that the spectra before the eclipse are generally characterized by prominent emission features and those after the eclipse start to show absorption features. At the eclipse, the smaller temperature gradient makes the spectra closer to blackbody emission. The transition can be easily seen by the strong features of \ce{CH4} at 3--4 and 7--9 micron. Similar transitions are also found for \ce{H2O} around 11 micron and HCN around 14 micron. Only minor features of CO at 4--5 micron, owing to the pressure level where CO absorbs the most coincide with the isothermal part of the atmosphere. Nevertheless, the \ce{CH4}--CO conversion can potentially be inferred by tracking \ce{CH4} and \ce{H2O} (also see Figure \ref{fig:emission_swap} for spectral variation due to trace composition isolated from the thermal variation). Within this JWST cycle 1 window between 10 hours before and after the eclipse, \ce{CO2} falls back to the lower equilibrium composition and cannot be detected in the emission spectra.

\begin{figure}
   \centering
   \includegraphics[width=\columnwidth]{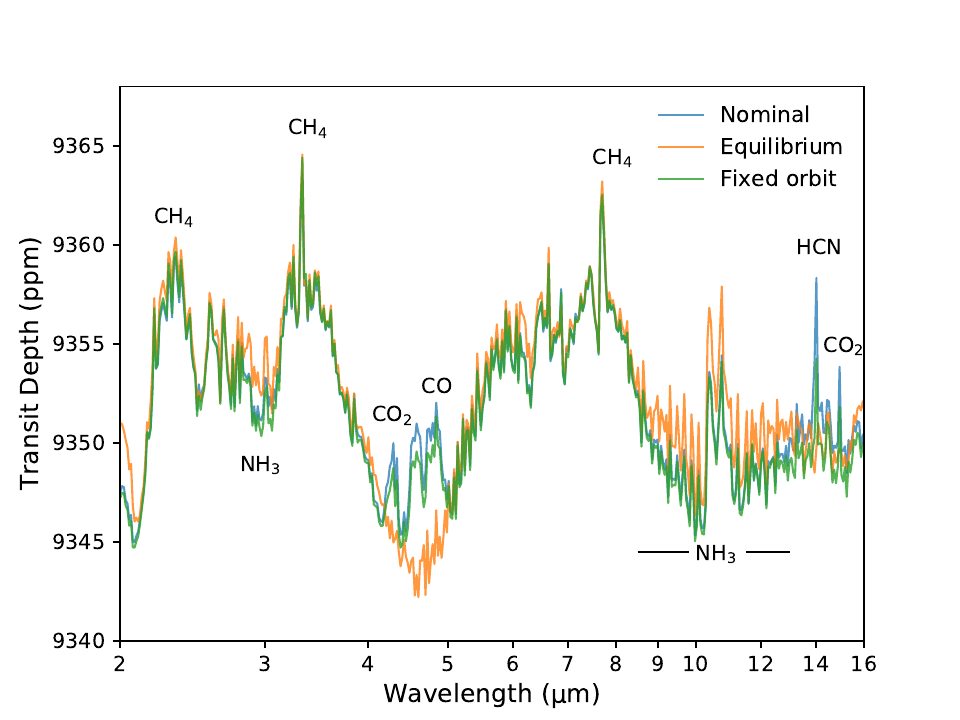}
      \caption{Transmission spectrum of HD 80606 b compared to that assuming chemical equilibrium and in steady state without the orbital motion.}
      \label{fig:transmission}
\end{figure}

Lastly, Figure \ref{fig:transmission} shows the theoretical transmission spectra of HD 80606 b, where the transit occurs about 5.6 days after periastron. CO, \ce{CO2}, and HCN exhibit significant disequilibrium features at 4--5 and 14 micron, respectively. The effects of orbit-induced quenching can also be seen by comparing the nominal kinetics model with the ``fixed-orbit'' model. Notably, leftover CO is carried over and contributes to the production of \ce{CO2} and HCN, where the feature of HCN at 14 micron shows the most pronounced differences. Since \ce{CO2} reached a peak value in the upper atmosphere during the transit, we suggest future transit observations to look for \ce{CO2} to put our photochemical modeling to the test. Although SO and \ce{SO2} also reached highest abundances before the transit, \ce{SO2} does not reach high enough abundances to be detected in all models we explored\footnote{The opacity of the more abundant SO is not included due to lack of data.}. 





\section{Discussion}\label{sec:discussion}
\subsection{Uncertainties in sulfur kinetics}
Considerable uncertainties exist in sulfur kinetics, especially the reactions involving reduced sulfur species \citep{Zahnle2016,Tsai2021}. One important initial step in forming sulfur allotropes is to make \ce{S2} from SH or S. For the three-body recombination reaction that form \ce{S2} from S, \ce{S + S ->[M] S2}, we followed previous works and adopted the latest calculation from \cite{Du2008}. This rate coefficient by \cite{Du2008} is consistent with \cite{Fair1969} but about three orders of magnitude smaller than another experimental rate \cite{Nicholas1979}. Thus, we consider our S${_x}$ formation in this study as a conservative limit. Another unsettled reaction that might be more relevant for HD 80606 b is \ce{S + CO ->[M] OCS}, whose reaction rates at low temperatures spread across a few orders of magnitude in the literature. \cite{Tsai2021} found that this reaction is key to OCS formation and \ce{S2} production on directly imaged gas giants, while \cite{Ranjan20} have shown that it can influence the CO abundance in a \ce{N2}--\ce{CO2} atmosphere as well. We reiterate the need of further investigation to resolve the OCS recombination rate.  

\subsection{Photochemical Hazes}
\begin{figure}
\begin{center}
\includegraphics[width=\columnwidth]{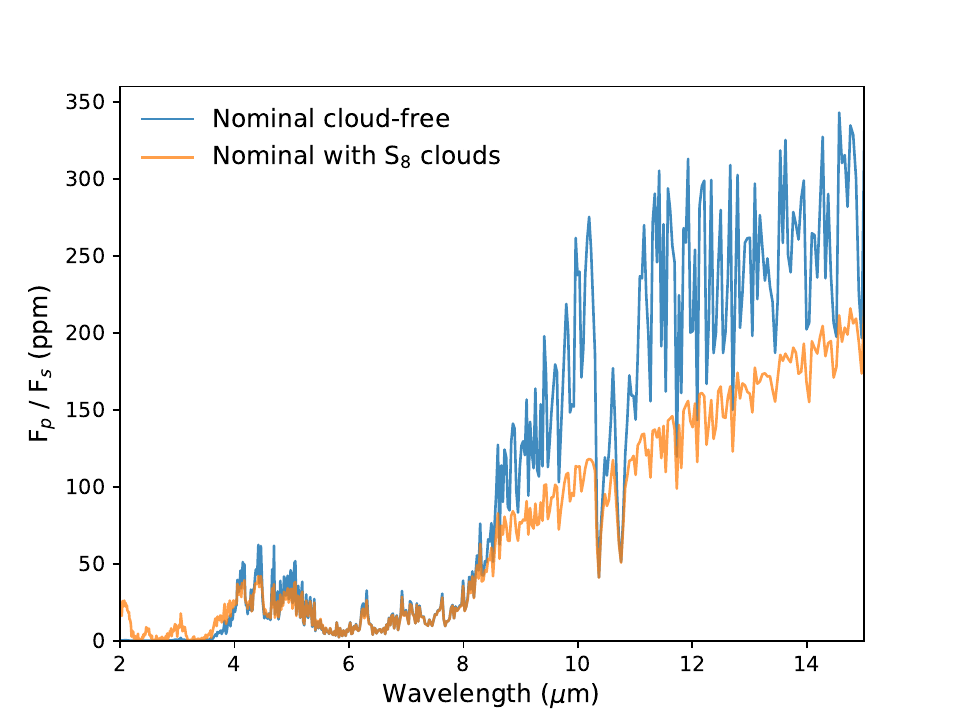}
\end{center}
\caption{Synthetic emission spectra of HD 80606 b at apoastron, with and without \ce{S8} aerosol opacity.}\label{fig:apo-S8clouds}
\end{figure}

The combination of \ce{CH4} carried over from the cold phase and the intense UV radiation seemingly entails an ideal condition for photochemical hazes near periastron. Indeed, our nominal models find copious \ce{C2H2} ($\sim$ 100 ppm) along with some hydrocarbon haze precursors, such as \ce{C6H6} (peaked around 0.01--0.1 ppm). If monomer formation and growth occur fast enough \citep{Seinfeld2016,Ohno2017}, organic hazes could affect the periastron passage. We will leave simulating the organic aerosols variations for future work (Ohno et al. in prep). On the other hand, \ce{S8} is produced outside of the periastron passage and condenses between 0.01 and 1 bar once the temperature cools down. \ce{S8} hazes scatter strongly and reduce the emission in the infrared. Figure \ref{fig:apo-S8clouds} illustrates 1 $\mu $m-sized \ce{S8} hazes substantially lower the emission flux. Our models suggest that organic hazes might be promoted during the periastron passage and deeper sulfur hazes could prevail a large portion of the orbit.


Additionally, \ce{Na2S} sodium sulfide could form via \ce{2Na + H2S -> Na2S(s) + H2}, where \ce{Na2S} particles condense out between 10 and 0.1 bar \citep{Lewis2017,Mayorga2021}. The \ce{Na2S} clouds can potentially suppress the emission flux but are evaporated and dissipated near periastron. In respect of the sulfur inventory, since sulfur is about 10 times more abundant than sodium in solar composition, it is unlikely that \ce{Na2S} would substantially deplete sulfur. 




\subsection{Future modeling and observational implications}\label{sec:disscusion}
We applied a 1D photochemical model based on averaged 3D GCM output to simulate the time-varying chemistry on HD 80606 b for this study. In reality, there will be radiative feedback from the compositional variation \citep[e.g.,][]{Drummond2016,Lavvas2021} and 3D circulation can complicate the global distribution of chemical species \citep[e.g.,][]{Mendonca2018a,Drummond2020,Steinrueck2021}. To assess the limit of our 1D photochemical model, we have compared the dynamical timescales and found that the horizontal transport timescale is generally shorter or comparable to the vertical mixing timescale. Since vertical quenching already occurred at 10 -- 100 bar, we expect horizontal transport simply lead to a uniform composition based on the vertically quenched abundance on the dayside. However, as shown in the study, photochemistry proceeds faster in the upper atmosphere and can have compelling interaction with the circulation. We encourage future work to treat the chemical feedback on the thermal structure consistently and/or utilize multidimensional photochemistry models \citep[e.g.,][]{Venot2020,Tsai2021b,Baeyens2022} to explore the global properties of the time-varying, eccentric systems. 


We have demonstrated that the quenched abundances of CO and \ce{CH4} before the \ce{CH4}-CO conversion kicks off around periastron are independent of the orbital phase. The measurement at the start of JWST cycle 1 observation should provide a baseline of the atmospheric properties. Our results suggest that [CO]/[\ce{CH4}] ratio is highly sensitive to metallicity and mildly sensitive to internal heat. Future spectral observations outside of the periastron passage of HD 80606 b can help us to place tighter constraints. Since \ce{CO2} is typically more sensitive to atmospheric metallicity, future transit observations will improve our ability to constrain atmospheric metallicity and validate the photochemical process.
 
We found that sulfur species are oxidized after periastron and SO and \ce{SO2} reach peak abundances near transit, while sulfur allotropes (S$_\textrm{x}$) exhibit a long-term growing sequence along the orbit. SH and \ce{S2} are strong absorbers in the optical and NUV (e.g., \ce{S2} emission has been reported after the impact of Shoemaker-Levy 9 on Jupiter \citep{Noll1995,Zahnle09}). \ce{S3} absorbs effectively near 400 nm and \ce{S4} has a weaker, broad-band absorption between 450 and 600 nm, whereas \ce{S8} has stretching mode features around 20 micron. We propose future observations sampling a wider coverage of the full phase to track the sequence of polysulfur species.

Finally, although we focused on HD 80606 b in this study, it would be illuminating to learn whether our findings can be generalized to other eccentric systems. While detailed atmosphere variation depends on the interplay between radiative and chemical processes, we expect photochemistry to play a main role in the time-dependent chemistry on other eccentric systems, given its short timescale. Exploring the key parameter space such as eccentricity and the irradiation temperature at periastron with coupling radiative transfer model and photochemical kinetics would be an ideal setting to address this question.

\section{Conclusions}\label{sec:sum}
We present an integrated modeling framework to study the atmospheric responses to the irradiation variation of the highly eccentric gas giant HD 80606 b. We applied a 3D GCM to simulate the fast-evolving climate dynamics and fed the dayside average atmosphere profiles to a time-dependent 1D photochemical model for the response of chemical compositions. We made predictions with synthetic spectra and demonstrate that our knowledge of atmospheric physics and chemistry will soon be put to the test with JWST observation.

We summarize our findings as follows,
\begin{itemize}
      \item The climate of HD 80606 b oscillates between the cold and hot states that are governed by distinct and well-studied dominating processes (day-night forcing vs. baroclinic instability).   
      
      \item Vertical quenching occurs in the deep atmosphere and is generally independent of the orbital position. Photochemistry from the intense UV radiation near periastron is the main driver of the compositional variation. 

      \item \ce{CH4} is partly converted to CO during the periastron passage and can be tracked by the JWST Cycle 1 GO observation. We have isolated the orbital effect in the model and showed that orbit-induced quenching plays an important role in this eccentric system. 
     
      \item We found \ce{CO2} reached a peak abundance near transit and could potentially be detected in transmission. Sulfur exhibits short-term response with oxidized species and long-term variations with sequence of allotropes cycle, where \ce{S8} is the main sulfur-bearing molecule above 0.1 bar away from periastron. 
      
     \item Our photochemical results suggest two classes of hazes as \ce{S8} condense out for most of the orbit while organic haze precursors are promoted near periastron.
\end{itemize}




\section*{Acknowledgements}
S.-M.T, M.S., and V.P. designed the project, where M.S. performed the GCMs, S.-M.T conducted photochemical models and wrote the manuscript. All authors discussed the results and contributed to the final manuscript. S.-M.T thanks Mark Hammond for the insightful discussion on the diagnosis of circulation and the exquisite college meals, X. Tan for a literature review, and J. Lyon for discussing sulfur chemistry. S.-M.T. acknowledges support from the European community through the ERC advanced grant EXOCONDENSE (\#740963; PI: R.T. Pierrehumbert). M.S. acknowledges support from
the Hubble Space Telescope through program GO-15246. Support
for this HST program was provided by NASA through a grant from
the Space Telescope Science Institute, which is operated by the
Association of Universities for Research in Astronomy, Inc., under
NASA contract NAS 5-26555.\\ \\
\section*{Data availability}
The model outputs underlying this work are available in the online supplementary material.
\bibliographystyle{aa}
\bibliography{master_bib}

\appendix
\onecolumn
\section{The compositional
variation of the C-, O-, and N- bearing species along the full orbit}
In this appendix, we show the variation of the abundances of several important carbon, oxygen, and nitrogen species along the full orbit to reiterate that the main chemical response occur during the short periastron passage while the composition of these species remain invariant for most of the orbit.
\begin{figure*}
   \centering
    \includegraphics[width=0.495\columnwidth]{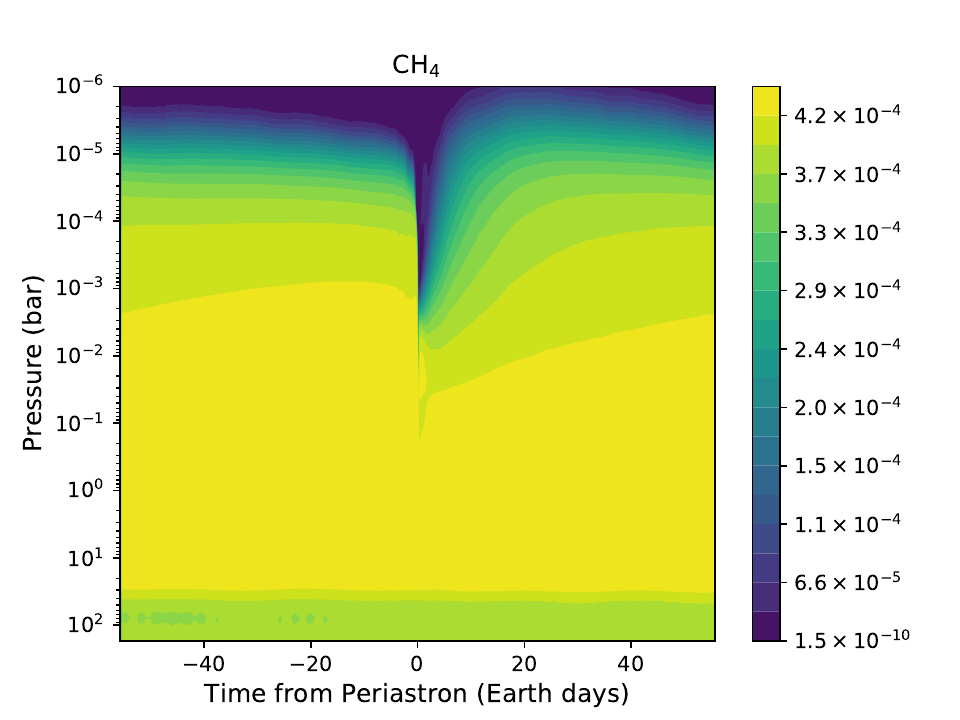}
    \includegraphics[width=0.495\columnwidth]{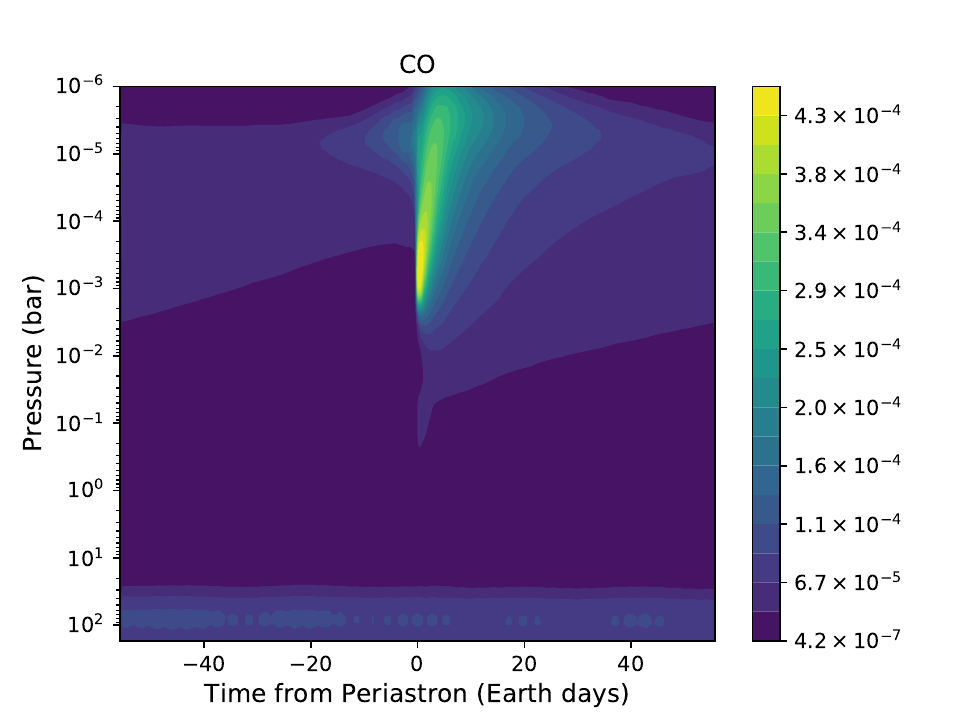}
    \includegraphics[width=0.495\columnwidth]{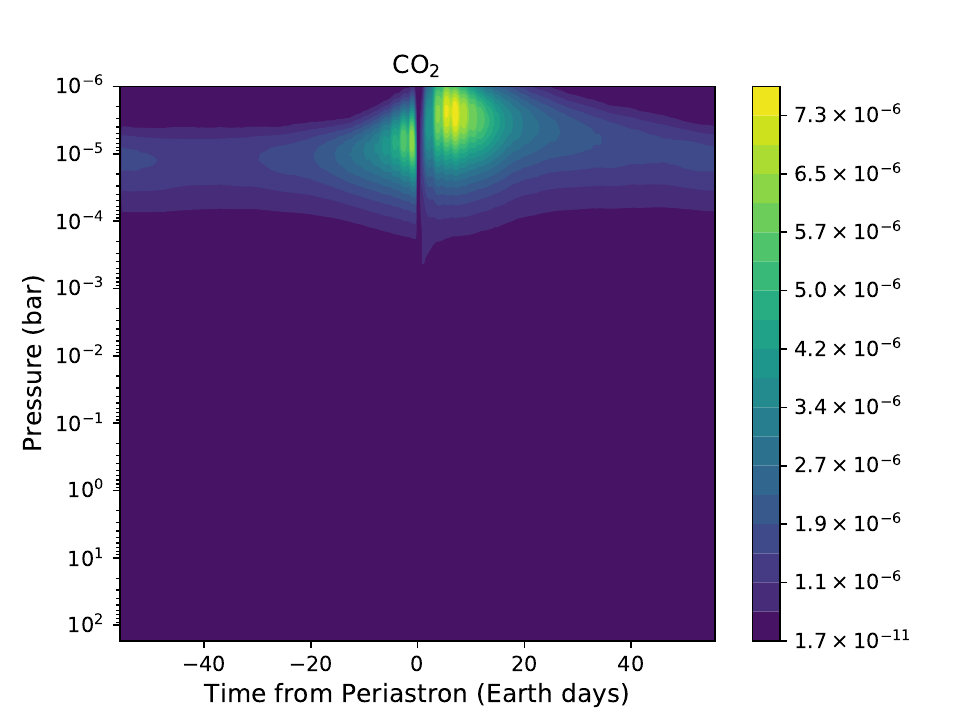}
    \includegraphics[width=0.495\columnwidth]{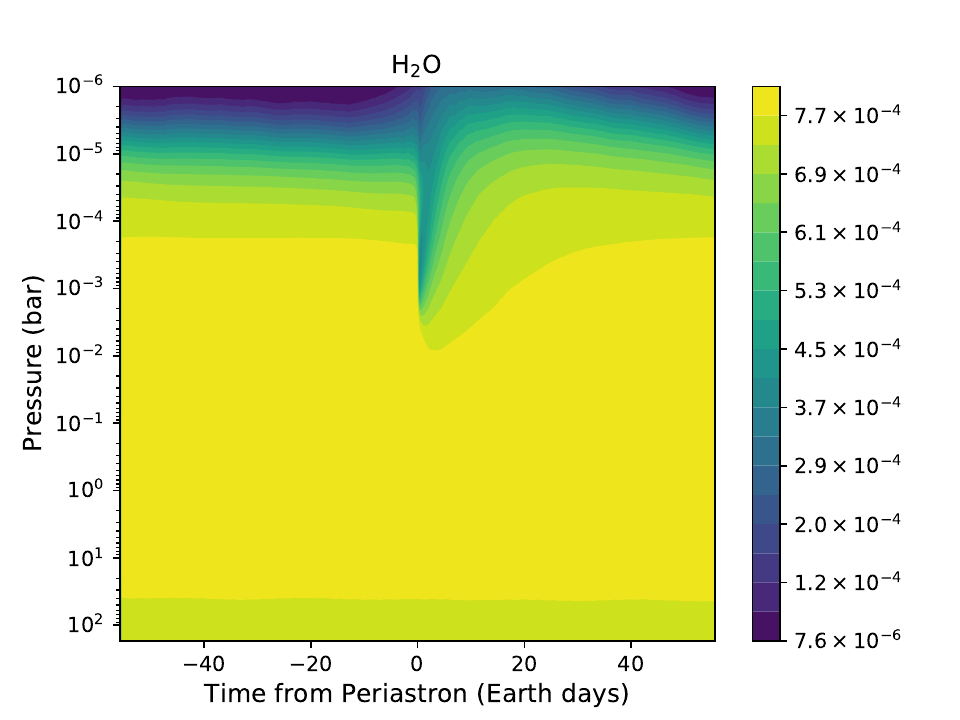}
    \includegraphics[width=0.495\columnwidth]{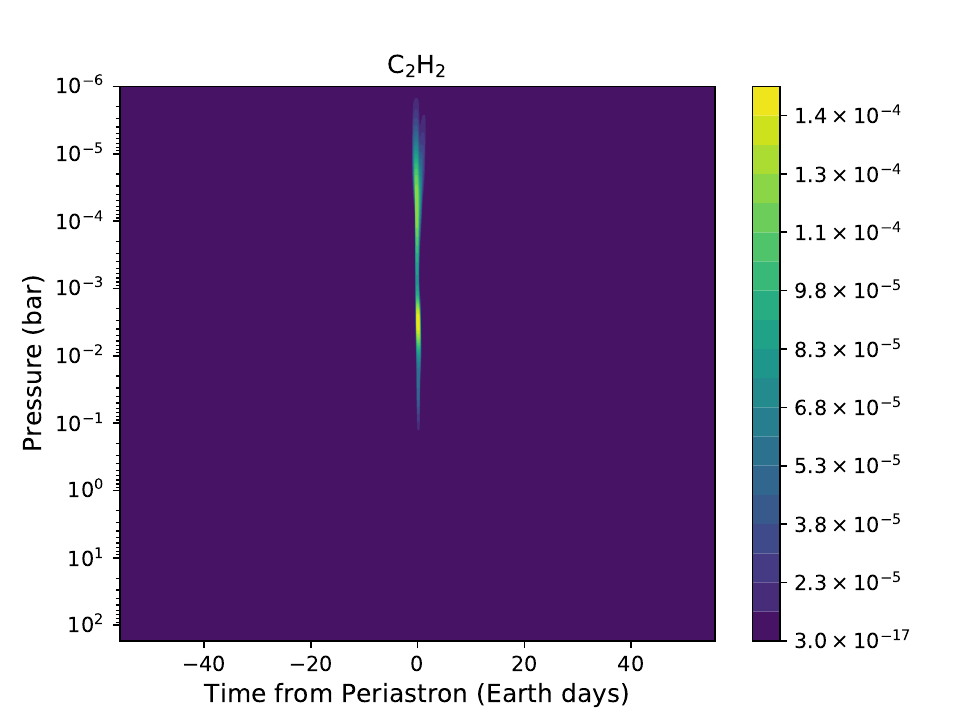}
    \includegraphics[width=0.495\columnwidth]{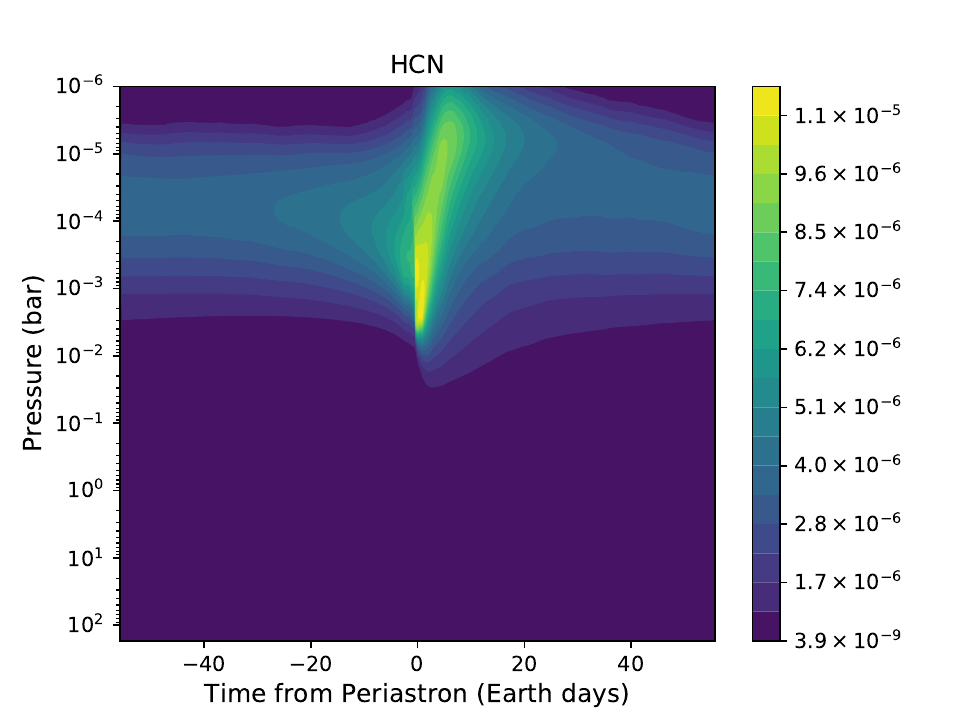}
      \caption{The dayside volume mixing ratios for several species of interest as a function of pressure and time (zero at periastron), like Figure \ref{fig:peri-contour} but showing the compositional variation across the full orbit.}
         \label{fig:full-contour} 
\end{figure*}
\clearpage
\section{Sensitivity to metallicity, internal heat, and vertical mixing}
Here, we present the composition distributions at apoastron and periastron when we vary the atmospheric metallicity, internal heat, and vertical mixing. 
\begin{figure}
   \centering
   \includegraphics[width=0.495\columnwidth]{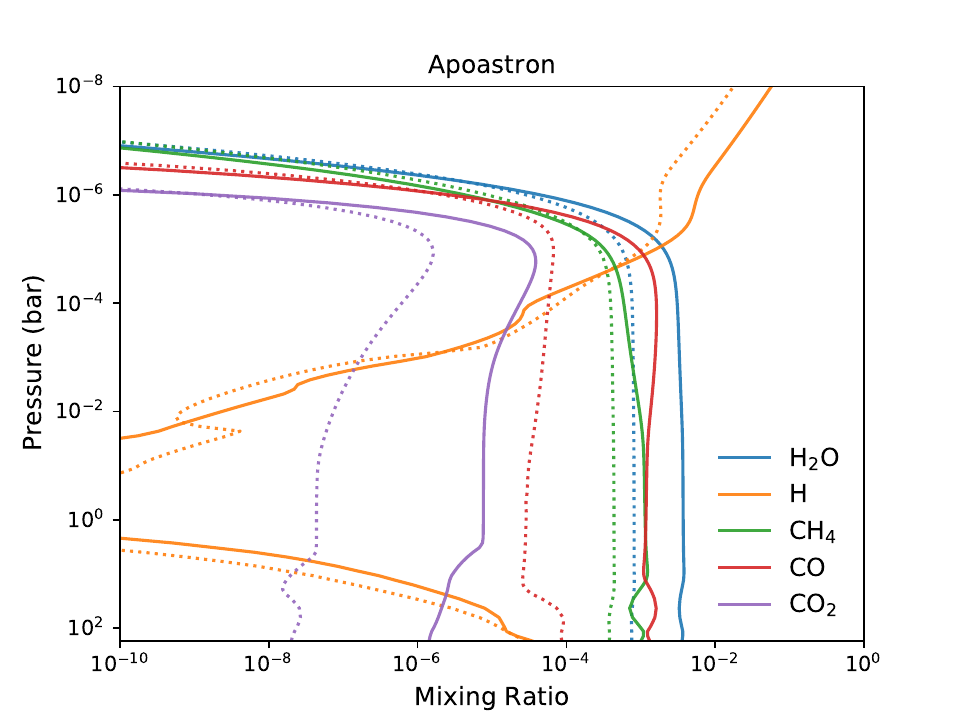}
   \includegraphics[width=0.495\columnwidth]{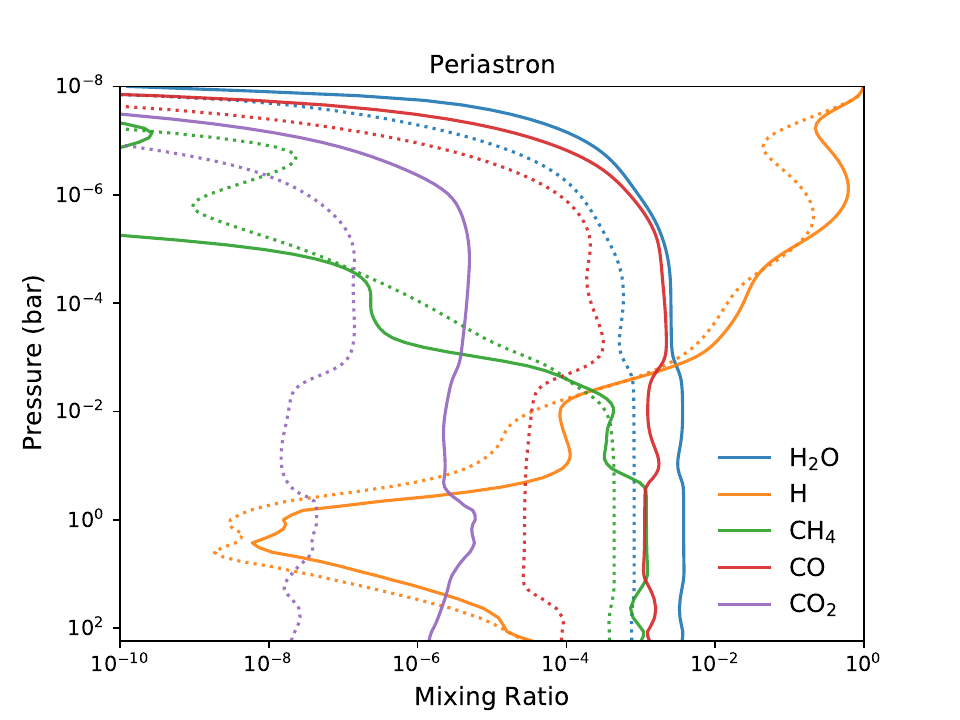}
   \includegraphics[width=0.495\columnwidth]{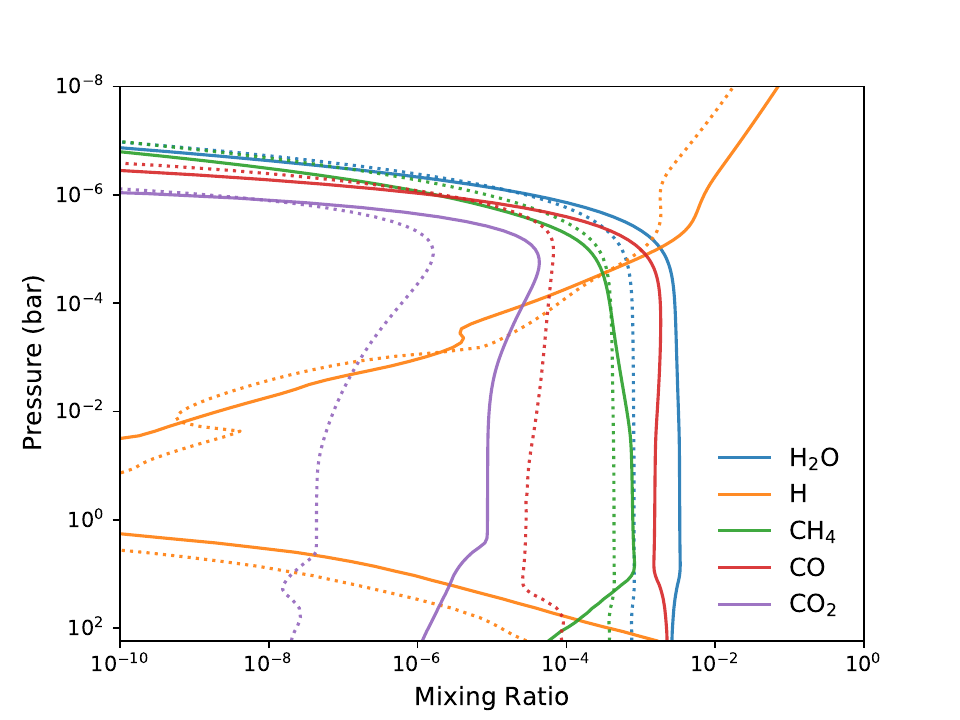}
   \includegraphics[width=0.495\columnwidth]{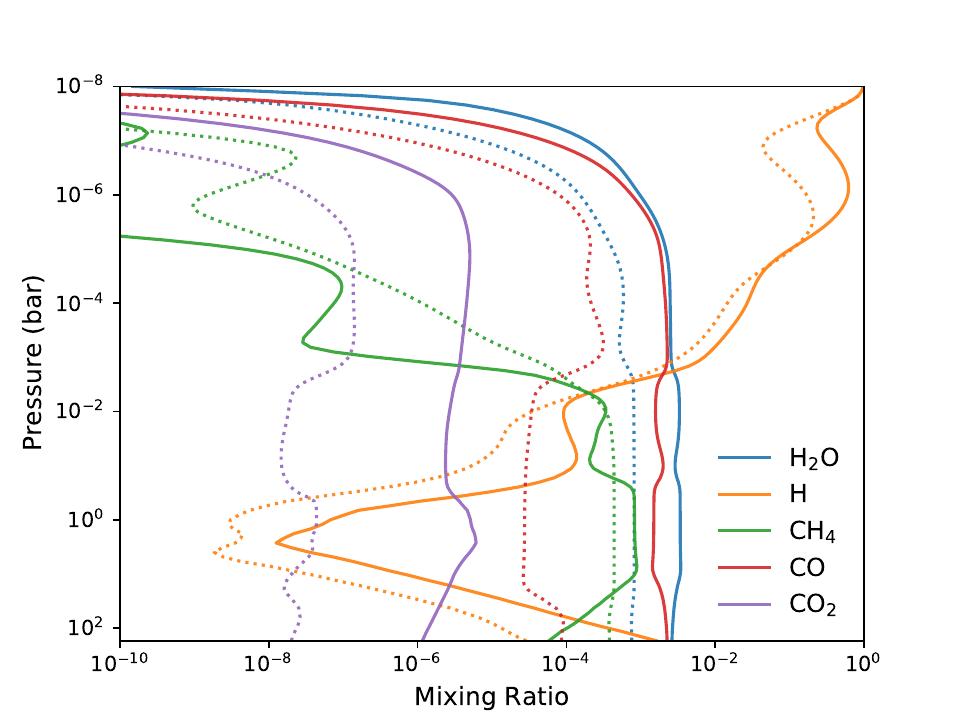}
      \caption{Composition profiles (solid lines) of HD 80606 b in our 5$\times$ solar metallicity, $T_{\textrm{int}}$ = 100 K model (top) and 5$\times$ solar metallicity, $T_{\textrm{int}}$ = 400 K model (bottom). The left panels are for apoastron and the right panels are for periastron. The nominal model with solar metallicity and $T_{\textrm{int}}$ = 100 K is shown in dotted lines for comparison.}
      \label{fig:1D-apo-peri-metallicity}
\end{figure}

\begin{figure*}
   \centering
   \includegraphics[width=0.495\columnwidth]{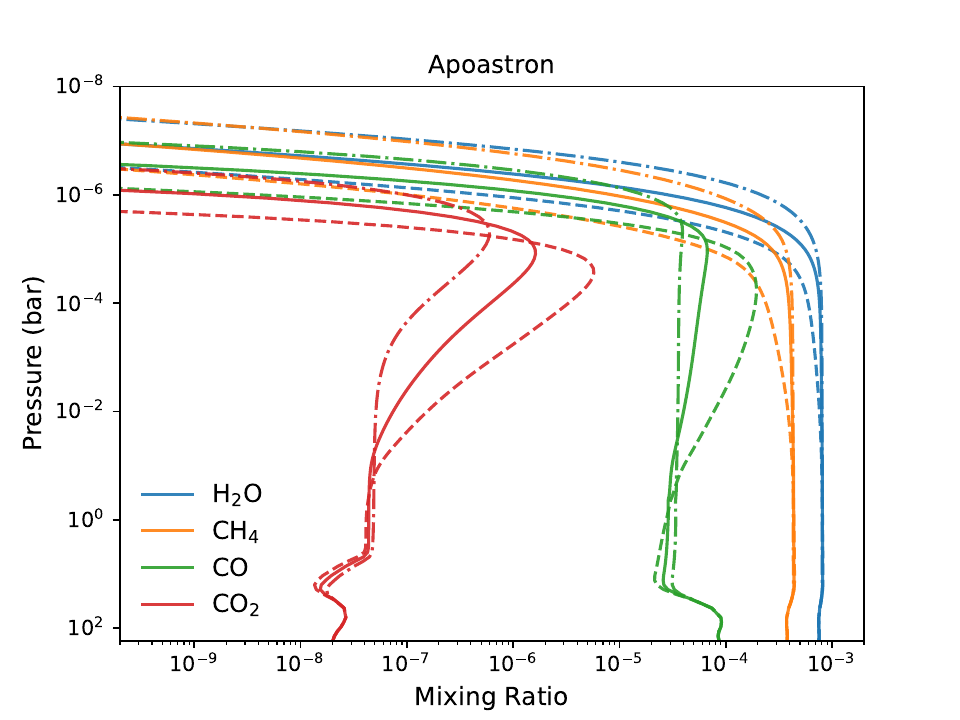}
   \includegraphics[width=0.495\columnwidth]{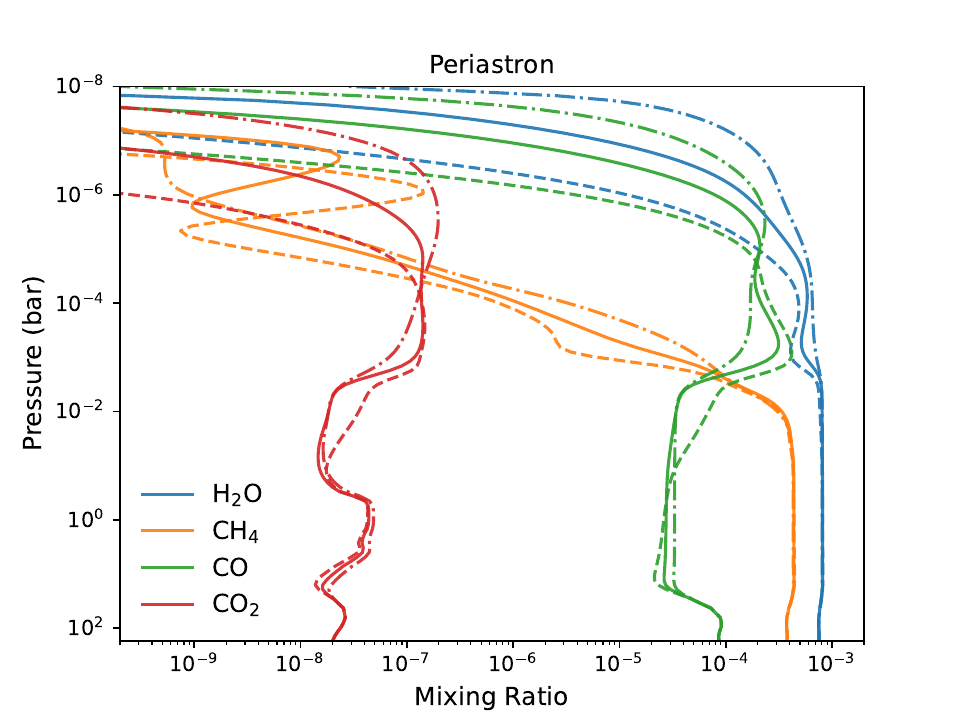}
      \caption{Composition profiles of HD 80606 b in our nominal model (solid), with 3 times (dashed-dotted) and 1/3 times (dashed) eddy diffusion coefficients.} 
      \label{fig:1D-Kzz}
\end{figure*}

\clearpage
\section{Spectral observability of compositional variation isolated from thermal variation}
\begin{figure}
\centering
\includegraphics[width=0.95\columnwidth]{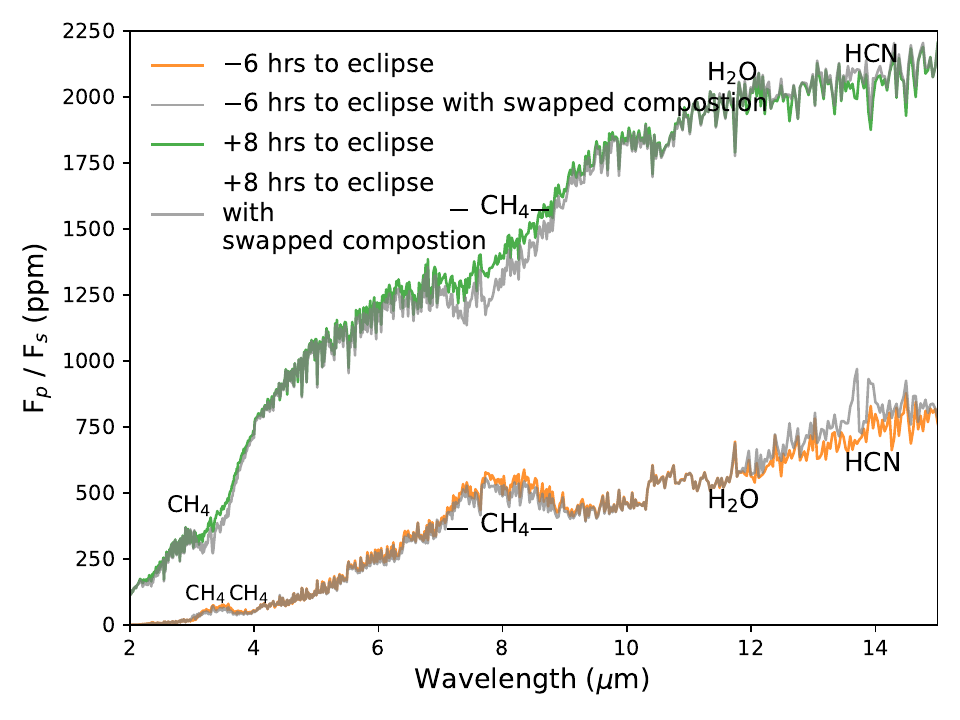}
  \caption{Emission spectra of HD 80606 b six hours before the eclipse and eight hours after the eclipse, similar to Figure \ref{fig:emission_time}, but here we show the spectra with swapped composition in grey (temperature profile -6 hr before the eclipse with composition +8hr after the eclipse and vice versa) to demonstrate the spectral changes in chemical composition isolated from thermal changes.}
      \label{fig:emission_swap}
\end{figure}

\label{lastpage}
\end{document}